\documentclass[times, 11pt]{article}    

\usepackage{times}
\usepackage{color,hcolor}
\usepackage{graphicx}
\usepackage{amsmath}
\usepackage{amssymb,amsthm}
\usepackage{subfigure}

\newcommand{\ket}[1]{|{#1} \rangle}

\newcommand{\bra}[1]{{\langle {#1}|}}

\newcommand{\fet}[1]{| {#1} \rangle^{\!\text{\tiny F}}}
\newcommand{\fra}[1]{{{}^{\,\text{\tiny F}}\!\langle {#1}|}}

\newcommand{\calu}{{\cal U}}

\newcommand{\mapm}[1]{\stackrel{#1}{\longrightarrow}}
\newcommand{\maps}[1]{\stackrel{\text{#1}}{\longrightarrow}}

\newtheorem{theorem}{Theorem}
\newtheorem{definition}{Definition}
\newtheorem{proposition}[theorem]{Proposition}

\begin{document}

\title{Quantum-Space Attacks}

\author{Ran Gelles \hskip 2cm  Tal Mor  \\      
\\
Technion - Israel Institute of Technology \\
Computer Science Department   \\
{\tt \{gelles, talmo\}@cs.technion.ac.il}
}

\maketitle

\begin{abstract}
Theoretical quantum key distribution (QKD) protocols
commonly rely on the use of qubits (quantum bits).
In reality, however, due to practical limitations,
the legitimate users are forced to employ a larger quantum
(Hilbert) space,
say a quhexit (quantum six-dimensional) space, or even
a much larger quantum Hilbert space.
Various specific attacks exploit
of these limitations. Although security can still be proved in some
very special cases,
a general framework that
considers such realistic QKD protocols,
{\em as well as} attacks on such protocols,
is still missing.

We describe a general method of attacking realistic QKD
protocols, which we call the `quantum-space attack'.
The description is based on assessing
the enlarged quantum space actually used by a protocol,
the `quantum space of the protocol'.
We demonstrate these new methods by
classifying various (known) recent
attacks against several QKD schemes,
and by analyzing a novel attack
on interferometry-based QKD.

\end{abstract}



\section{Introduction}\label{sec:Introduction}
%
%
Quantum cryptography has brought us new ways of
exchanging a secret key between two
users (known as Alice and Bob).
The security of such Quantum Key Distribution (QKD) methods
is based on a very basic rule of nature and quantum mechanics---the
``no-cloning'' principle.
The first QKD protocol was suggested in a seminal paper by
Bennett and Brassard~\cite{BB84} in 1984, and is now known
as BB84.
During recent years many security analyses were published
\cite{Y95, Mayers01, BBBMR-STOC, ShorPres00, BBBMR, Gisin05}
which proved the information-theoretical security of
the BB84 scheme
against the most general attack by an unlimited adversary (known as Eve),
who has full control over the quantum channel\footnote{All QKD protocols
assume that Alice and Bob also use an insecure, yet unjammable, classical channel.}.
Those security proofs are
limited as they always
consider a theoretical QKD that uses perfect qubits.
Although these security proofs do take errors into account,
and the protocols use error correction and privacy amplification (to
compensate for these errors and for reducing any partial knowledge that Eve
might have),
in general, they
avoid security issues that arise from the implementation
of qubits in the {\em real world}.

%
%
A pivotal paper by Brassard,
L\"utkenhaus, Mor, and Sanders~\cite{BLMS-ec00,BLMS00}
presented the ``Photon Number Splitting (PNS) attack'' and
exposed a security flaw in experimental and practical QKD:
One must take into account the fact that Alice
does not generate perfect qubits (2 basis-states of a single photon),
but, instead, generates states that reside in an enlarged
Hilbert space (we call it ``quantum space'' here), of six dimensions.
The reason for that discrepancy in the size of the used quantum
space is that each
electromagnetic pulse that Alice generates
contains (in addition to the two dimensions spanned by the single-photon states)
also a vacuum state and three 2-photon states,
and these are extremely useful to the eavesdropper.
That paper proved that, in contrast to what was assumed in
previous papers,
Eve can make use of the enlarged space,
and get a lot of
information on the secret key, sometimes even full information,
without inducing any noise.
Many attacks on the practical protocols then
followed (e.g.,~\cite{H03,HLP04,GLLP04,NSG05,HM05,GFKZR06}),
based on
extensions of the quantum spaces,
exploring various additional security flaws; other papers~\cite{H03,SARG04,W05}
suggested possible ways to overcome such attacks.
On the one hand,
several security proofs, considering specific imperfections,
were given for the BB84 protocol \cite{GLLP04, ILM2007}.
Yet on the other hand, it is
generally impossible now to prove the security of
a practical protocol, since
{\em a general framework} that
considers such realistic QKD protocols,
{\em and} the possible attacks on such protocols,
is still missing.

%
%
We show that
the PNS attack, and actually all attacks directed at
the channel, are various special cases
of a general attack that we define here,
the {\em Quantum-Space Attack} (QSA).
The QSA generalizes existing attacks and also offers novel attacks.
The QSA is based on the fact that the ``qubits'' manipulated in
the QKD protocol actually reside in a larger Hilbert space, and this enlarged
space {\em can be assessed}. Although this enlarged space is
not fully accessible to the legitimate users,
they can still analyze it, and learn what a fully powerful eavesdropper
can do.
We believe that this assessment of the
enlarged ``quantum space of the protocol''
is a vital step on the way to proving
or disproving the unconditional security of practical QKD schemes.
We focus on schemes in which the quantum communication is
uni-directional, namely, from Alice's laboratory (lab) to Bob's lab.
We consider
an adversary that can attack all the quantum states
that come out of Alice's lab,
and all the quantum states that go into Bob's lab.

%
%
The paper is organized as follows:
Definitions of the
quantum spaces involved in the
realization of a protocol, and of the
``quantum space of the protocol'',
are presented
and discussed in Section~\ref{sec:QSoP}.
The ``quantum-space attack'' is defined and discussed in Section~\ref{sec:QSA}.
Using the general framework when the information carriers are
photons is discussed in
Section~\ref{sec:QSAphotonicWorld}. 
Next, in Section~\ref{sec:knownQSA} we show 
that the best known attacks on practical QKD
are special cases of the QSA.
Section~\ref{sec:InterferoBB84} demonstrates and analyzes
a novel QSA on an interferometric implementation of the BB84
and the six-state QKD protocols.
Last, we discuss a few subtleties and open problems for future research
in Section~\ref{sec:conclusion}.

We would like to emphasize that our (crypt)analysis presents
the difficulty of proving unconditional
security for practical QKD setups,
yet also provides an important (probably even vital)
step in that direction.


\section{The Quantum Space of the Protocol}
\label{sec:QSoP}

%
%
The Quantum Space Attack (QSA) is the most general attack
on the quantum channel that connects Alice to Bob. It
can be applied to any realistic
QKD protocol, yet here we focus on uni-directional schemes
and on implementations of the BB84 protocol and the six-state protocol.
We need to have a proper model of the protocol
in order to understand the Hilbert space
that an unlimited Eve can attack.
This space has never been analyzed before except for specific cases.
Our main finding is a proper description of this space,
which allows, for the first time,
defining the most general eavesdropping attack on the channel.
We start with a model of a practical ``qubit'', continue with understanding
the spaces used by Alice and Bob, and end by defining the relevant space,
the {\em Quantum Space of the Protocol}
(QSoP), used by Eve to attack the protocol.
The attacks on the QSoP are what we call {\em Quantum-Space Attacks}.


\subsection{Alice's realistic space}
\label{sec:HA}
%
%
In most QKD protocols, Alice sends Bob qubits, namely,
states of
2 dimensional quantum spaces ($H_2$).
A realistic view should take into account
any deviation from theory, caused by Alice's equipment.
For example, Alice might encode the qubit via a polarized photon:
$\ket{0_z}$ via a photon polarized horizontally, and
$\ket{1_z}$ polarized vertically. This can be written using
Fock notation\footnote{States written using the Fock notation
$\fet{\cdot}$ are called Fock states, see Section~\ref{sec:QSAphotonicWorld}.}
 as $\fet{n_{h},n_v}$ where $n_h$ ($n_v$) represents
the number of horizontal (vertical) photons; then
$\ket{0_z} \equiv \fet{1,0}$ and $\ket{1_z} \equiv \fet{0,1}$.
When Alice's photon is lost within her equipment
(or during the transmission), Bob gets the
state $\fet{0,0}$,
so that Alice's realistic space becomes $H_3$. Alice might send
multiple photons and then $H^A$ is of higher dimension, see
Section~\ref{sec:AliceRealPhotonic}.
\begin{definition}
\label{def:HA}
{\bf Alice's realistic space, $H^A$,}
is the minimal space containing the actual quantum states
sent by Alice to Bob during the QKD protocol.
\end{definition}

%
%
In the BB84 protocol, Alice sends qubits in two\footnote{The six-state scheme
uses the three conjugate bases of the qubit space;
namely, also
 $\ket{0_y}=\left (\ket{0_z}+i\ket{1_z}\right)/\sqrt{2}$, etc.}
fixed conjugate bases.
Theoretically, Alice randomly chooses a basis and a bit value
and sends the chosen bit encoded in the appropriate chosen basis as
a state in $H_2$ (e.g.\ $\ket{0_z}$,$\ket{1_z}$,
 $\ket{0_x}=\left (\ket{0_z}+\ket{1_z}\right)/\sqrt{2}$, and
 $\ket{1_x}= \left (\ket{0_z}-\ket{1_z}\right)/\sqrt{2}$).
To a better approximation, the states sent by Alice
are four different states
$\ket{\psi_i}_A$ ($i= 1,2,3,4$)
in her realistic space $H^A$, spanned by these
four states. This space $H^A$ is
of dimension $|H^A|$, commonly between 2 and 4,
depending on the specific implementation.
As practical instruments often diverse from theory,
Alice might send quite different states. As an extreme example,
see the {\em tagging attack} (Section~\ref{sec:TagAsQSA}),
which is based on the fact that Alice's space could
contain more than just these four theoretical states, so that
$|H^A| > 4$ is possible.


\subsection{Extension of Alice's space}
\label{sec:ExtendAliceSpace}

%
%
Bob commonly receives one of several possible states
$\ket{\psi_i}_{A}$  sent by Alice, and measures it. The most
general measurement Bob can perform is to add an ancilla, perform
a unitary transformation on the joint system,  perform a
complete measurement,
and potentially
``forget''\footnote{By the term
``forget'' we mean that Bob's detection is unable
to distinguish between several measured states.}
some of the outcomes\footnote{This entire process can be described
in a compact way  by using a POVM~\cite{Peres93}.}. However, once
Alice's space is larger than $H_2$, the extra dimensions provided
by Alice could be used by Bob for his measurement,
{\em instead of} adding an ancilla.
%
%
Interestingly, by his measurement Bob might be {\em extending}
the space vulnerable to Eve's attack well beyond $H^A$.
This is possible since in many cases the realistic space, $H^A$,
is embedded inside a larger space $M$.
\begin{definition}
\label{def:M}
The space $M$ is the space in which
$H^A$ is embedded, $H^A \subseteq M$.
The space $M$ is the actual space available for Alice and an Eavesdropper.
\end{definition}

%
%
Due to the presence of an eavesdropper,
Bob's choice whether to add an ancilla or
to use the extended space $M$
is vital for security analysis.
In the first case the ancilla
is added by Bob, inside his lab,
while in the second it is controlled by Alice,
transferred through the quantum channel
and exposed to Eve's deeds.
Eve might attack the extended space $M$,
and thus have a different
effect on Bob, considering his measurement method.

%
%
For example,
suppose Alice sends two non-orthogonal states of a qubit,
$\theta_0 = {\cos\theta \choose \sin\theta}$
and $\theta_1 = {\cos\theta \choose -\sin\theta}$, with
a fixed and known angle $0 \ge \theta \ge 45^{\circ}$.
Bob would like to distinguish between them,
while allowing inconclusive results
sometimes, but no errors~\cite{Peres88}.
Bob can add the ancilla $\ket{0}_{{ Anc}} \equiv {1 \choose 0}_{Anc}$
and perform the following
transformation~$\calu$:
\begin{multline}
 \ket{0}_{{ Anc}} \otimes {\cos\theta \choose \pm \sin\theta}
=
\left ( \begin{array}{c} \cos\theta \\ \pm\sin\theta \\0 \\0
\end{array} \right )
\mapm{\calu}
\left ( \begin{array}{c} \sin\theta \\ \pm\sin\theta \\
\sqrt{\cos 2\theta} \\ 0 \end{array} \right )
\\
 = \sqrt{2}\sin\theta
\ket{0}_{{ Anc}} \otimes {1/\sqrt{2} \choose \pm 1/\sqrt{2}} +
  \sqrt{\cos 2\theta} \ket{1}_{{ Anc}} \otimes {1 \choose 0}
\end{multline}
where
$\ket{1}_{{ Anc}} \equiv {0 \choose 1}_{Anc}$.
This operation leads to a conclusive result with probability
$2\sin^2\theta$ (when the measured ancilla is $\ket{0}_{{ Anc}}$),
and inconclusive result otherwise. It is simple to see that the
same measurement can be done, {\em without the use of an ancilla},
if the states $\theta_0$ and $\theta_1$ are embedded at
Alice's lab in a larger
space $M$, e.g.\  $M = H_3$, using Bob's transformation
\begin{equation}\label{eqn:3dimMeasurmetn}
\left ( \begin{array}{c} \cos\theta \\ \pm\sin\theta \\ 0 \end{array} \right )
\mapm{\calu}
\left ( \begin{array}{c} \sin\theta \\ \pm\sin\theta \\
       \sqrt{\cos 2\theta} \end{array} \right ) \text{.}
\end{equation}
In the general case, the space $M$ might be very large,
even infinite. Bob might use only
parts of it, for his measurements.

%
%
A complication in performing security analysis is due to
Bob's option to {\em both} use an ancilla and extend
the space used by Alice.
Our analysis in the following sections starts with the
space extension only (Sections \ref{sec:HB}--\ref{sec:HP}),
and later on deals with the general case (Sections
\ref{sec:HB+anc}--\ref{sec:HP+anc}).


\subsection{Bob's space, without an ancilla}
\label{sec:HB}
%
%
Let us formulate the spaces involved in the protocol, as described
above. Assume Alice uses the space $H^A$ according to
Definition~\ref{def:HA},
which is embedded in a
(potentially larger) space $M$.
Ideally, in the BB84 protocol,
Bob would like to measure just the states in $H^A$, but
in practice he usually can not do so.
Each one of Alice's states $\ket{\psi_i}_{ A}$ is transformed
by Bob's equipment into some pure\footnote{The case in which
Bob transposes the state into a mixed state is a special case
of the analysis done in Section~\ref{sec:HB+anc}.
For the notion of mixed states or quantum mixture see~\cite{NC00,Peres93}.}
state $\ket{\psi_i}_{ M} \in M$. The
space which is spanned by those states contains all the
information about Alice's states $\{ \ket{\psi_i}_{ A} \}$.

%
%
More important, Bob might be measuring un-needed
subspaces of $M$ which Alice's states do not span.
For instance, examine the case where Bob uses detectors to measure
the Fock states $\fet{1,0}$ and $\fet{0,1}$. Bob is usually able
to distinguish a loss (the state $\fet{0,0}$) or an error (e.g.\ $\fet{1,1}$,
one horizontal photon and one vertical photon),
from the two desired states,
but he cannot distinguish between other states containing multiple photons.
This means that Bob measures a much larger subspace of
the entire space $M$, but (inevitably) interprets
outcomes outside $H^A$ as legitimate states;
e.g.\ the states $\fet{2,0}$, $\fet{3,0}$, etc. are (mistakenly)
interpreted as $\fet{1,0}$.
See further discussion in Section~\ref{sec:PhotonicExtension}.

%
%
We denote Bob's setup
(beam splitters, phase shifters, etc.) by the unitary
operation $\calu_B$, followed by a measurement; all these
operations are operating on the space $M$ (or parts of it).
Bob might have several different setups
(e.g.\ a different setup for the $z$-basis and for the $x$-basis).
Let $\mathbf{U}$ be the set of unitary transformations
in all Bob's setups.
%
%
\begin{definition} \label{def:HB}
{\bf [This definition is Temporary.]}
Given a specific setup-transformation $\calu_j \in \mathbf{U}$, let
$H^{B_{j}} \subseteq M$ be the subsystem actually measured by Bob,
having $K$ basis states $\{ \ket{\phi_k}_{B_j} \}_{k= 0 \ldots K-1}$.
The set of {\bf Bob's Measured Spaces}
is the set $\{ H^{B_{j}} \}_{j=0 \ldots J-1}$
of $J= |\mathbf{U}|$ spaces.
\end{definition}

We have already seen that Bob might be measuring un-needed dimensions.
On the other hand he might not measure
certain subspaces of $M$, even when Alice's state might reach there.
In either case, the deviation
is  commonly due to limitations of Bob's equipment.


\subsection{The quantum space of the protocol, without an ancilla}
\label{sec:HP}
%
%
The ``quantum space of the protocol'' (QSoP) is in fact Alice's {\em extended}
space, taking into consideration its {\em extensions}
due to Bob's measurements.
The security analysis of a protocol
depends on the space $H^{B^{-1}}$ defined below.
\begin{definition}\label{def:HB-1}
{\bf [This definition is Temporary.]}
{\bf The reversed space $H^{B^{-1}}$} is the Hilbert space
spanned by the states $\calu_j^{-1} (\ket{\phi_k}_{B_j})$,
for each possible setup $\calu_j \in \mathbf U$,
and for each basis state
$\ket{\phi_k}_{B_j}$
of the appropriate $H^{B_j} \subseteq M$.
\end{definition}
The Space $H^{B^{-1}}$
usually resides in a larger space than $H^A$.
For instance, using photons, the ideal space $H^A$ consists of
two modes with 2 basis states,
see Section~\ref{sec:QSAphotonicWorld}.
Now $H^{B^{-1}}$ could have an infinite space in each mode,
but also could have more modes.

%
%
In order to derive the quantum space of the protocol we need to
define the way Alice's space is extended according to $H^{B^{-1}}$,
for this simple case where Bob does not add an ancilla.
In this case, the space $H^{B^{-1}}$ simply extends Alice's space to
yield the QSoP via $H^P = H^A + H^{B^{-1}}$.
Formally speaking
%
%
\begin{definition}\label{def:HP}
{\bf [This definition is Temporary.]}
{\bf The Quantum Space of the Protocol}, $H^P$,
is the space spanned by the basis states of the space $H^A$
and the basis states of the space $H^{B^{-1}}$.
\end{definition}
If Alice's realistic space
is fully measured by Bob's detection process,
then $H^A$ is a subspace of $H^{B^{-1}}$, hence
$H^P = H^{B^{-1}}$.


\subsection{Bob's space (general case)}
\label{sec:HB+anc}
%
%
In the general case, one must consider Bob's option to add an ancilla
during his measurement process.
This addition causes a considerable
difficulty in analyzing a protocol, however
it is often an inherent part of the protocol,
and can not be avoided.
We denote the added ancilla as the state $\ket{0}_{B'}$
that resides in the space $H^{B'}$.
\begin{definition}\label{def:M+anc}
$M'$ is the space that includes the physical space
used by Alice as defined in Definition~\ref{def:M},
in addition to Bob's ancilla, $M' = M \otimes H^{B'}$.
\end{definition}
Bob measures a subspace of the space $M'$, so the
(permanent)
definitions of his measured spaces $H^{B_j}$ and the reversed space
$H^{B^{-1}}$ should be modified accordingly.
\begin{definition} \label{def:HB+anc}
Given a specific setup-transformation $\calu_j \in \mathbf{U}$ let
$H^{B_{j}} \subseteq M'$ be the subsystem actually measured by Bob,
having $K$ basis states $\{ \ket{\phi_k}_{B_j} \}_{k= 0 \ldots K-1}$.
The set of {\bf Bob's Measured Spaces},
is the set $\{ H^{B_{j}} \}_{j=0 \ldots J-1}$
of $J= |\mathbf{U}|$ spaces.
\end{definition}

\subsection{The quantum space of the protocol (general case)}
\label{sec:HP+anc}

The quantum space of the protocol is still Alice's {\em extended}
space, while considering its {\em extensions}
due to Bob's measurements. Yet, the added ancilla makes things
much more complex.
The security analysis of a protocol
depends now {\em not} on the space $H^{B^{-1}}$ defined below,
but on a (potentially {\em much larger})
space obtained from it by tracing-out Bob's ancilla.
As before, we first define the reversed space.
\begin{definition}\label{def:HB-1+anc}
{\bf The reversed space $H^{B^{-1}}$} is the Hilbert space
spanned by the states $\calu_j^{-1} (\ket{\phi_k}_{B_j})$,
for each possible setup $\calu_j \in \mathbf U$,
and for each basis state
$\ket{\phi_k}_{B_j}$
of the appropriate $H^{B_j} \subseteq M'$.
\end{definition}

%
%
Once a basis state of one of
Bob's measured spaces $\ket{\phi_k}_{B_j}$
is reversed by $\calu_j^{-1}$ we result with a state that
might, partially, reside in Bob's ancillary space $H^{B'}$.
Since Eve has no access to this space\footnote
{Giving this space to Eve
(for getting an upper bound on her information),
might be easier to analyze, but
is usually not possible since it would
give her too much power, making the protocol insecure.}
it must be traced-out (separated out), for deriving the QSoP.
Let us redefine the QSoP given the addition of the ancilla:
\begin{definition}\label{def:HP+anc}
{\bf The Quantum Space of the Protocol, $H^P$}, is the space
spanned by {\bf (a)} the basis states of the space $H^A$; and {\bf (b)}
the states $\mathrm{Tr}_{Bob}[ \calu_j^{-1} (\ket{\phi_k}_{B_j})]$,
(namely, after tracing out Bob),
for each possible setup $\calu_j \in \mathbf U$, and
for each basis state $\ket{\phi_k}_{B_j}$ of the appropriate
space $H^{B_j}$.
\end{definition}
%
%
Whenever $\calu_B$ entangles Bob's ancilla with the system sent from Alice,
tracing out Bob's ancilla after performing $\calu_B^{-1}$
might cause an increase of the QSoP to the dimension of
Bob's ancillary space.
For instance, assume Alice's state is embedded in an $n$-qubit space
to which Bob adds an ancilla of $n$-qubits and performs a unitary
transformation $\calu$, such that for one state measured
by Bob, $\ket{\Psi}_{B} \mapm{\calu^{-1}}
\frac{1}{2^{n/2}}\sum_{k=0}^{2^n-1}
\ket{k}_{P}\ket{k}_{B'}$.
Tracing out Bob from this state yields the
maximally mixed state
$\rho_{P} = \frac{1}{2^n}\sum_{k=0}^{2^n-1} \ket{k}\bra{k}$,
so that in this example
the whole $n$-qubits space is spanned.


\section{The Quantum Space Attack}
\label{sec:QSA}

\subsection{Eavesdropping on qubits}

%
%
When Alice and Bob use qubits, in theoretical QKD, Eve can attack
the protocol in many
ways.  In her simplest attack, the so-called ``measure-resend attack'',
Eve performs any measurement (of her choice) on the qubit,
and accordingly decides what to send to Bob.

%
%
A generalization of that attack is the ``translucent attack'',
in which Eve attaches an ancilla, in an initial state $\ket{0}_E$ (and in any
dimension she likes), and entangles the ancilla and Alice's qubit, using
$\ket{0}_E \ket{i}_A \rightarrow \sum_{j=0}^{1} \ket{E_{ij}}_E \ket{j}_A$
where $\ket{i}_A$ is a basis for Alice's qubit, and Eve's states after
the unitary transformation are $\ket{E_{ij}}_E$.
Using this transformation one can define the most general
``individual-particle attack''~\cite{EHPP94,FGGNP97},
and also the most general
``collective attack''~\cite{BM97a,BBBGM02}.
In the individual-particle attack
Eve delays the measurement of her ancilla till after learning
anything she can about the qubit
(e.g., its basis), while in the collective attack Eve delays her measurements
further till she learns anything she can
about {\em all} the qubits (e.g., how the final key is generated from the
obtained string of shared bits), so she attacks directly the {\em final key}.

%
%
The most general attack that Eve could perform on the channel
is to attack all those qubits
transmitted from Alice to Bob, using {\em one} large ancilla.
This is the ``joint attack''.
Security, in case Eve tries to learn a maximal information on the final key,
was proven in~\cite{Y95,Mayers01,BBBMR-STOC,ShorPres00,BBBMR} via various methods.
The attack's unitary transformation is written as before, but with
$i$ a binary string of $n$ bits, and so is $j$,
$  \ket{0}_E\ket{i}_A\rightarrow \sum_{j=0}^{2^n-1}
\ket{E_{ij}}_E\ket{j}$.

\subsection{Eavesdropping on the quantum space of the protocol}

%
%
By replacing the qubit space $H_2$ by Alice's realistic ``qubit'' in the space
$H^A$, and by defining Eve's attack on the entire space of the protocol
$H^P$, we can generalize each of the known attacks on theoretical QKD
to a ``quantum space attack'' (QSA).
We can easily define now Eve's most general
{\it individual-transmission QSA}
on a realistic ``qubit'',
which generalizes the individual-particle attack earlier described.
Eve prepares an ancilla in a state $\ket{0}_{E}$,
and attaches it to Alice's state, but actually her ancilla is
now attached to the entire QSoP.
Eve performs a unitary
transformation $\calu_E$ on the joint state.
If Eve's attack is only on $H^A$, we write
the resulting transformation on any basis state
of $H^A$,
$\ket{i}_A$, as
$\ket{0}_E\ket{i}_A \rightarrow \sum_j \ket{E_{ij}}_{E}\ket{j}_{A}$,
where the sum is over the dimension of $H^A$.
The Photon-Number-Splitting attack (see Section~\ref{sec:PNSasQSA})
 is an example for such an attack.
The most general
individual-transmission QSA is based on a translucent
QSA on the QSoP,
\begin{equation}
\label{eqn:TranslucentQSA}
\ket{0}_E\ket{i}_P \rightarrow \sum_j \ket{E_{ij}}_{E}\ket{j}_P\text{,}
\end{equation}
where the sum is over the dimension of $H^P$.
The subsystem in $H^P$ is then sent to Bob
while the rest (the subsystem $H^E$) is kept by Eve.
We write
the transformation on any basis state
of $H^P$, $\ket{i}_P$, but
note that it is sufficient to define the transformation on the different
states in $H^A$, namely
for all states of the form $\ket{i}_A$, since other states of the QSoP are
never sent by Alice (any other additional subsystem of  the QSoP
is necessarily at a known state when it enters Eve's transformation).

%
%
Attacks that are more general than the {\it individual
transmission QSA}, the
{\it collective QSA} and the {\it joint QSA},
can now be defined
accordingly.
In the most general collective QSA, Eve performs the above
translucent QSA
on many (say, $n$) realistic ``qubits''
(potentially a different attack on each one, if she
likes), waits till she gets all data regarding the generation
of the final key,
and she then measures all the ancillas together, to obtain the optimal
information on the final key or the final secret.
The most general attack that Eve could perform on the channel
is to attack all those realistic ``qubits''
transmitted from Alice to Bob, using
{\em one} large ancilla. This is the
``joint QSA''.
The attack's unitary transformation is written as before, but with
$i$ a string of $n$ digits rather than a single digit
(digits of the relevant dimension of $H^P$),
and so is $j$,
\begin{equation}
\label{eqn:JointQSA}
 \ket{0}_E\ket{i}_{P^{\otimes n}} \rightarrow \sum_{j=0}^{{|H^P|}^n-1}
 \ket{E_{ij}}_E\ket{j}_{P^{\otimes n}}\text{.}
\end{equation}
Eve measures the ancilla, after learning all classical information,
to obtain the optimal
information on the final key or the final secret.
As before, it is sufficient to define the transformation on
the different input states from $(H^A)^{\otimes n}$.

%
%
We would like to emphasize several issues:
1.-- When analyzing specific attacks,
or when trying to obtain a limited
security result,
it is always legitimate to restrict the analysis to
the relevant (smaller) subspace of the QSoP, for simplicity,
e.g., to $H^A$, or to $H^{B^{-1}}$, etc.
2.-- Any bi-directional protocol will have a much more complicated QSoP,
thus it might be extremely difficult to analyze any type of
QSA (even the simplest ones) on such protocols.
This remark is especially important since
bi-directional protocols play a very important
role in QKD, since they appear in many
interesting protocols such as the
plug-and-play~\cite{MHHTG97},
the ping-pong~\cite{BF02}, and the classical Bob~\cite{BKM07} protocols.
Specifically they provided (via the plug-and-play)
the only commerical QKD so far~\cite{idQ, magiQ}.
3.-- It is well known that the collective or joint
attack is only finished after Eve gets all
quantum and classical information, since she delays her measurements
till then~\cite{BM97a, BBBGM02, BBBMR-STOC, Mayers01, BBBMR}; if she expects
more information, she better wait and attack the final secret rather
than the final key;  it is important to notice that if the key will be used
to encode quantum information (say, qubits) then the quantum-space of the
protocol will require a modification, potentially a major one;
It is interesting to study if this new notion of QSoP has an influence on
analysis of such usage of the key as done (for the ideal qubits)
in~\cite{BHLMO05}.


\section{Photonic Quantum Space Attacks}
\label{sec:QSAphotonicWorld}

\subsection{Photons as quantum-information carriers}
\label{sec:PhotonAsQuantumSystem}

%
%
Since most of the practical QKD experiments and products are done
using photons, in this section we demonstrate
our QSoP and QSA definitions and methods via photons.
Our analysis uses the Fock-Space\footnote{A description of the
Fock space and Fock notations can be found in
various quantum optic books, e.g.\ \cite{SZ97}.} notations
for describing photonic quantum spaces.
For clarity, states written using the Fock notation are denoted
with the  superscript `{\tiny F}', e.g.\ $\fet{0}$,
$\fet{3}$, and
$\fet{0,3,1}$.

%
%
A photon can not be treated as a quantum system
in a straightforward way. For instance, unlike dust particles or grains
of sand,
photons are indistinguishable particles, meaning that when a couple of
photons are interacting, one cannot define the evolution of the specific
particle, but rather describe the whole system.

%
%
Let us examine a cavity, for instance.
It can contain photons of specific wavelengthes
($\lambda_1$, $\lambda_2$, etc.) and
the energy of a photon of wavelength $\lambda$ is
directly proportional to $1/\lambda$.
While one cannot distinguish between photons of the same wavelength, one can
distinguish between photons of different wavelengths.
Therefore, it is convenient to define distinguishable
``photonic modes'', such that each wavelength corresponds to a specific mode
(so a mode inside a cavity can be denoted by its wavelength),
and then count the number of photons in each mode.
If a single photon in a specific mode carries some unit of energy, then $n$
such photons of the same wavelength carry $n$ times that energy.
If the cavity is at its ground (minimal) energy level,
we say that there are
``no photons'' in the cavity and denote the state as $\fet{0}$---the vacuum
state.
The convention is to denote only those modes that
are potentially populated, so
if we can find $n$ photons in one mode, and no photons in any other mode,
we write, $\fet{n}$.
If two modes are populated by $n_a$ and $n_b$ photons, and all
other modes are surely empty,
we write $\fet{n_a,n_b}$ (or $\fet{m,n}_{ab}$).
When there is no danger of confusion, and the number
of photons per mode is small (smaller than ten), we just write
$\fet{mn}$ for $m$ photons in one mode and $n$ in the other.
In addition to its wavelength, a photon also has a property called
polarization, and a basis for
that property is, for instance, the horizontal and vertical polarizations
mentioned earlier.  Thus, two modes (in a cavity) can also
have the same energy,
but different polarizations.

%
%
Outside a cavity photons travel with the speed of light,
say from Alice to Bob,
yet modes can still be described, e.g., by using ``pulses'' of light~\cite{BLPS90}.
The modes can then be distinguished by different directions of the light
beams (or by different paths),
or by the timing of pulses (these modes are denoted by non-overlapping
time-bins), or by orthogonal
polarizations.

%
%

%
%
A proper description of a photonic qubit is
commonly based on using two modes `$a$' and `$b$' which are
populated by exactly a single photon,
namely, a photon in mode $a$, so the state is $\fet{10}_{ab}$,
or a photon in mode $b$,
so the state is $\fet{01}_{ab}$.
However, a quantum space that consists of a
single given photonic mode `$a$' is not restricted to a single photon,
and can be populated by any number of photons.
A basis for this space is $\{\fet{n}_a\}$ with $n \ge 0$, so that
the quantum space is infinitely large, $H_\infty$.
Theoretically, a general state in this space is can be written
as the superposition $\sum_{n=0}^\infty c_{n}\fet{n}_a$,
with $\sum_n|c_n|^2=1$, $c_n \in \mathbb{C}$.
Similarly, a quantum space that consists of two photonic modes
has the basis states $\fet{n_a,n_b}$, for $n_a , n_b \ge 0$ and
a general state is of the form
$\sum_{n_a , n_b = 0}^\infty c_{n_a,n_b}\fet{n_a,n_b}$
with $\sum_{n_a , n_b = 0}^\infty |c_{n_a,n_b}|^2 = 1$,
$c_{n_a,n_b} \in \mathbb{C}$.
This quantum space is described as a tensor product of two ``systems''
$H_\infty \otimes H_\infty$.
%

%
%
Using {\em exactly} two photons in two different
(and orthogonal) modes assists in clarifying
the difference between photons and dust particles (or grains of sand):
Due to the indistiguishability of photons, only 3 different
states can exist (instead of 4): $\fet{20}_{ab}$, $\fet{02}_{ab}$ and
$\fet{11}_{ab}$.
The last state has one photon in  mode `$a$' and another photon 
in `$b$', however, exchanging the photons is meaningless
since one can never tell one photon from another.

%
%
A realistic model
of a photon source (in a specific mode)
is of a coherent pulse (a Poissonian distribution)
\[
\ket{\alpha} = e^{- \frac{|\alpha|^2}{2}}
\sum^{\infty}_{n=0}\frac{\alpha^n}{\sqrt{n!}}\ket{n}
\]
including terms
that describe the possibility of emitting any number $n$ of photons.
As the number of photons increases beyond some number,
the probability decreases, so it
is common to neglect the higher orders.
In QKD, experimentalists commonly use a ``weak'' coherent state (such that
$|\alpha|\ll 1$) and then terms with $n\ge3$ can usually be neglected.
There is also a lot of research about
sources that emit (to a good approximation) single photons,
and then, again,
terms with $n\ge3$ can usually be neglected.

%
%


\subsection{Alice's realistic photonic space}
\label{sec:AliceRealPhotonic}
%
%
While the theoretical qubit lives in $H_2$,
a realistic view defines the space actually used by Alice
to be much larger.
The possibility to emit empty pulses
increases Alice's realistic space into $H_3$,
due to the vacuum state $\fet{00}_{ab}$.
When Alice sends a qubit using two modes,
using a weak coherent state (or a ``single-photon'' source),
her realistic space, $H^A$,
is embedded in $H_{\infty}\otimes H_{\infty}$.
Terms containing more than two photons can be neglected,
so these are excluded from Alice's space $H^A$.
The appropriate realistic quantum space of Alice, $H^A$,
is now a quhexit: the six-dimensional space
spanned by
$\chi^6 =\{\fet{00}$, $\fet{10}$,
$\fet{00}$, $\fet{11}$, $\fet{20}$, $\fet{02} \}$.
The PNS attack 
demonstrated in Section~\ref{sec:PNSasQSA},
is based on attacking
this 6 dimensional space $H^A$.
Note also that terms with more than two photons still appear in $M$,
and thus could potentially appear in the QSoP (and then used by Eve).

%
%
At times, Alice's realistic space is even larger, due to extra modes
that are sent through the channel, and are not meant to be
a part of the protocol. 
These extra modes might severely compromise the security of the protocol,
since they might carry some vital information about the protocol.
A specific QSA based on that flaw is the 
``tagging attack'' (Section~\ref{sec:TagAsQSA}).
Note that even if Alice uses exactly two modes, the quantum space
$M$ where $H^A$ is embedded, certainly contains other modes as well.

\subsection{Extensions of the photonic space; the QSoP}
\label{sec:PhotonicExtension}

%
%
Let us discuss Bob's measurement of photonic spaces.
There are (mainly) two types of detectors that can be used.
The common detector can not distinguish
a single photon from more than one photon
(these kind of detectors are known as {\em threshold detectors}).
The Hilbert space where Bob's measurement
is defined is infinite\footnote{
In practice, that space is as large as Eve
might wish it to be.
We can ignore the case where Eve uses too
many photons so that the detector
could burn due to the high energy,
since it is not in Eve's interest.
Thus, in some of the analyses below we replace
$\infty$ by some large number $L$.},
since a click in the detector tells Bob that the
number of photons occupying the mode is ``not zero'' i.e.\
the detector clicks when $\fet{n}$ is detected,
for $n \ge 1$.
This means that Bob
measures the state $\fet{0}$, or he measures
 $\fet{1}$, $\fet{2}$, $\ldots$ but then ``forgets'' 
how many photons were detected.
Bob might severely compromise the security,
since he inevitably interprets a
measurement of a state containing multiple photons
as the ``legal'' state that contains only a single photon.
An attack based on a similar limitation is the
``Trojan-Pony'' attack
described below, in Section~\ref{subsec:Trojan}.
In order to avoid false interpretations of the photon number
reaching the detector, Bob could use an enhanced type of detector
known as the {\em photon-number resolving detector}
or a {\em counter} (which is still under development).
This device distinguishes a single photon from $n \ge 2$ photons, hence any
eavesdropping attempt that generates multi-photon states can
potentially be noticed by Bob.
A much enhanced security
can be achieved now,
although the QSoP is infinite also in this case,
due to identifying correctly the legitimate state $\fet{1}$, from
various legitimate states.

%
%
The number of modes in the QSoP
depends on Bob's detectors as well.
Bob commonly increases the number of measured
modes by ``opening''
his detector for more time-bin modes or more frequency modes.
For instance, suppose Bob is using a detector
whose detection time-window is quite larger than the width of the
pulse used in the protocol, since he does not know when exactly Alice's
pulse might arrive.
The result is an extension of the space used by Alice,
so that the QSoP includes the subspace
of $M$ that contains all these measured modes.
When a single detector is used to measure
more than one mode {\em without distinguishing them}, the
impact on the security might be severe, see the
``Fake state'' attack (Section~\ref{sec:FakeAsQSA}).

In addition to the known attacks described in the following subsection,
a new QSA is analyzed in Section~\ref{sec:InterferoBB84}, where
we examine the more general case of QSA, in which
Bob adds an ancilla during the process.

%

\section{Known Attacks as Quantum-Space Attacks}\label{sec:knownQSA}

All known attacks can be considered as special cases of the
Quantum-Space Attack. In this section we show a description of
several such attacks using QSA terms. For each and every attack we
briefly describe the specific protocol used, the quantum space of
the protocol, and a realization of the attack as a QSA.


\subsection{The photon number splitting attack~\cite{BLMS00}}
\label{sec:PNSasQSA}
{\bf The Protocol.} Consider a BB84 protocol, where Alice uses a
``weak pulse'' laser to send photons in two  modes corresponding
to the vertical and horizontal polarizations when using the $z$
basis (the diagonal polarizations then relate to using the $x$
basis). Bob uses a device called a Pockel cell to rotate the
polarization (by $45^\circ$) for measuring  the $x$ basis, or
performs no rotation if measuring  the $z$ basis. The measurement
of the state is then done using two detectors and a ``polarization
beam splitter'' that passes the first mode to one detector and the
second mode to the other detector (for a survey of
polarization-based QKD experiments, see~\cite{GRTZ02, DLH06}).

{\bf The Quantum Space of the Protocol.} Every pulse sent by Alice 
is in one of four states, each in a superposition of the 6
orthogonal states $\chi^6 = \{\fet{00}$, $\fet{10}$,
$\fet{01}$,$\fet{11}$,$\fet{20}$, $\fet{02}\}$, where the space
used by Alice is $H^A = H_6$. Bob uses two setups, $\calu_{B_z}=
I$ for the $z$ basis, and ${\calu_{B_x}}$ for the $x$ basis, which
is more complex and described in Appendix~\ref{app:polU}.

The detectors used by Bob cannot distinguish between modes having
single photon and multiple photons. Each one of his two detectors
measures the basis elements $\{ \fet{n} \}$ for $n\ge0$ (of the
specific mode directed to that specific detector), where Bob
interprets the states $\{ \fet{n} \}$ with $n > 1$ as measuring
the state $\fet{1}$ of the same mode. Bob's measured space $H^{B}$
is thus infinite and spanned by the states $\{ \fet{mn} \}$ for
$m,n \ge 0$. The QSoP $H^P$ is equal to $H^{B_z}$ ($=H^{B_x}$)
since performing $\calu^{-1}$ does not change the dimensionality of
the spanned space (in both setups).

{\bf The Attack.}
Eve measures the number of photons in the pulse, using
non-demolition measurement. If she finds that the number of
photons is $\ge 1$, she blocks the pulse and generates a loss. In
the case she finds that the pulse consists of 2 photons, she
splits one photon out of the pulse and sends it to Bob, keeping
the other photon until the bases are revealed, thus getting full
information of the key-bit. Eve sends the eavesdropped qubits to
Bob via a lossless channel so that Bob will not notice the
enhanced loss-rate. As is common in experimental QKD, Bob is
willing to accept a high loss-rate (he does not count losses as
errors), since most of Alice's pulses are empty. See the precise
mathematical description of this attack in
Appendix~\ref{App:MathPNS}.


\subsection{The tagging attack (based on~\cite{GLLP04})}\label{sec:TagAsQSA}
{\bf The Protocol.} Consider a BB84 QKD protocol in which  Alice
sends an enlarged state rather than a qubit. This state contains,
besides the information qubit, a {\em tag} giving Eve some
information about the bit. The tag can, for example, tell Eve the
basis being used by Alice. For a potentially realistic example,
let the tag be an additional qutrit indicating if Alice used the
$x$-basis, or the $z$-basis, or whether the basis is {\em
unknown}: whenever Alice switches basis, a single photon comes out
of her lab prior to the qubit-carrying pulse, telling the basis,
say using the states $\fet{10}_{\textit{tag}}$ and
$\fet{01}_{\textit{tag}}$, and when there is no change of basis,
what comes out prior to the qubit is just the vacuum
$\fet{00}_{\textit{tag}}$.

{\bf The Quantum Space of the Protocol.} In this example, Alice is
using the space $H^A = H_2 \otimes H_{\textit{tag}} = H_2 \otimes
H_3$. Bob, unaware of the enlarged space used by Alice, expects
and receives only the subspace $H_2$. We assume that Bob
ideally measures this space with a single setup $\calu_B = I$,
therefore $H^{B} = H_2$. Since Bob's setup does not change the
space, $H^{B^{-1}} = H_2$ as well. However, the tag is of a much
use to Eve, and indeed the QSoP following Definition~\ref{def:HP},
defined  to be $H^P = H_2 \otimes H_{\textit{tag}}$.

{\bf The Attack.} Eve uses the tag in order to
retrieve information about the qubit without inducing error (e.g.\
via cloning the qubit in the proper basis).
The attack is then an intercept-resend QSA. We mention that this
attack is very similar to a side-channel cryptanalysis of classic
cryptosystems.

\vskip 12 pt
{\bf A Short Summery.} It can be seen that the PNS attack
described above is actually a special case of the tagging attack,
where the {\it tag} in that case is in fact another copy of the
transmitted qubit. This copy is kept by Eve until the bases are
revealed, then it can be measured so the the key-bit value is
exposed with certainty. Both those QSA attacks are based on the
fact that Alice (realistic) space is larger than the theoretical
one. Although in the PNS example, the QSoP is further extended due
to Bob's measurement, the attack is not based on that extension but
on the fact that $H^A$ is larger than $H_2$. In the following
attacks  Bob's measurements cause the enlargement of the QSoP,
allowing Eve to exploit the larger QSoP for her attack.


\subsection{The Trojan-pony attack \cite{GLLP04, HLP04}}
\label{subsec:Trojan}
%
%
In Trojan-pony attacks Eve modifies the state sent to Bob in a way
that gives her information. In contrast to a ``Trojan-horse'' that
goes in-and-out of Bob's lab, the ``pony'' only goes in,
therefore, it is not considered an attack on the lab, but only on
the channel. We present here an interesting example~\cite{GLLP04}.

{\bf The Protocol.} Assume a polarization-encoded BB84 protocol,
in which Alice is ideal, namely, sending perfect qubits
($H^A=H_2$). However, Bob uses realistic threshold detectors that
suffer from losses and dark counts, and that cannot distinguish
between one photon and $k$ photons for $1<k<L$.
In order to be able to ``prove'' security, for a longer distance
of transmission Bob wants to keep the error-rate low although the
increase of dark counts' impact with the
distance~\cite{BLMS00}. Therefore, Bob assumes that Eve has no
control over dark counts, and whenever both detectors click, Alice
and Bob agree to consider it as {\em a loss} since it is outside
of Eve's control (i.e.\ the QSoP is falsely considered to be
$H_2$). Namely, they assume that {\em an error} occurs only when
Bob measures in the right basis, and only one detector clicks,
(which is the detector corresponding to the wrong bit-value).

%
%
{\bf The Quantum Space of the Protocol.} Same as in
Section~\ref{sec:PNSasQSA}, Bob's measured spaces $H^{B_z}$,
$H^{B_x}$, the reversed space $H^{B^{-1}}$ as well as the QSoP
$H^P$, are merely the spaces describing two modes (with up to $L$
photons), $H_L \otimes H_L$.
Bob's detectors cannot distinguish between receiving a
single-photon pulse from a multi-photon pulse, so his measurement
is properly described as a projection of the received state onto
the space containing $\{\fet{ij}\}$ followed by ``forgetting'' the
exact result, and keeping only one of three results:
``$\{10\}\equiv$ detector-1 clicks'', ``$\{01\}\equiv$ detector-2
clicks'', and else it is $\{00\}$, a ``loss''. In formal, {\em
generalized-measurements} language (called POVM,
see~\cite{Peres93,NC00}) these three possible results are written
as: $\{10\}\equiv \sum_{k=1}^{L-1} \fet{k0}\fra{k0}$,
$\{01\}\equiv \sum_{k=1}^{L-1} \fet{0k}\fra{0k}$, $\{00\}\equiv
\fet{00}\fra{00} + \sum_{k_1,k_2\  =1}^{L-1}
\fet{k_1k_2}\fra{k_1k_2}$, and their sum is the identity matrix.

%
%
{\bf The Attack.} Eve's attack is the following: (a) Randomly
choose a basis (b) Measure the arriving qubit in that specific
chosen basis (c) Send Bob $m$-photons identical to the measured
qubit, where $m \gg 1$. Obviously, when Eve chooses the same basis
as Alice and Bob then Bob measures the exact value sent by Alice,
and Eve gets full information. Otherwise, both of his detectors
click, implying a ``loss'', except for a negligible probability,
$\approx 2^{(-m+1)}$, thus Eve induces no errors.
%
%
The main observation of this measure-resend QSA is that treating a
count of more than a single photon as a loss, rather than as an
error, is usually not justified. A second conclusion is that
letting Bob use counters instead of threshold detectors (to
distinguish a single photon from multiple photons), together with
treating any count of more than one photon as an error, could be
vital for proving security against QSA.  The price is that dark
counts put severe restrictions on the distance to which
communication can still be considered secure, as suggested already
by~\cite{BLMS00}.


\subsection{The fake-state attack (based on~\cite{HM05,MAS06})}
\label{sec:FakeAsQSA}

{\bf The Protocol.} In this example, we examine a polarization
encoded BB84 protocol, and an ideal Alice ($H^A = H_2$). This time
Bob's detectors are imperfect so that their detection windows do
not fully overlap, meaning that there exist times in which one
detector is blocked (or it has a low efficiency), while the other
detector is still regularly active. Thus, if Eve can control the
precise timing of the pulse, she can control whether the photon
will be detected or lost. The setup is built four detectors and a
rotating mirror (since Bob does not want to spend money on a
Pockel cell (polarization rotator), he actually uses 2 fixed
different setups). Using the rotating mirror Bob sends the photon
into a detection setup for basis $z$ or a detection setup for
basis $x$. Suppose the two detection setups use slightly different
detectors, or slightly different delay lines, or slightly
different shutters, and Eve is aware of this (or had learnt it
during her past attacks on the system).
For simplicity, we model the non-overlapping detection windows, as
additional two modes, one slightly prior to Alice's intended mode
(the pulse), and one right after it.

{\bf The Quantum Space of the Protocol.} The original qubit is
sent in a specific time-bin $t_0$ (namely, $H^A=H_2$). The setup
$\calu_Z$ is a set of two detectors and a polarized beam splitter,
separating the horizontal and the vertical modes to the detectors,
where $\calu_x$ separate the diagonal modes into a set of two
(different) detectors. Let the detectors for one basis, say $z$,
be able to measure a pulse arriving at $t_0$ or $t_1$, while the
detectors for the other basis ($x$) measure pulses arriving at
$t_{-1}$ or $t_0$.

For simplicity, we degenerate the space to contain one or less
photons\footnote {As mentioned above, this is used for
non-security proof, and is not legitimate assumption for proving
unconditional security, where the three time-modes
 should be considered as $H_L \otimes H_L \otimes H_L$.},
so that $H^{B_z}$ is $H_5$, i.e. two possible time-bins consisting
each of two (polarization) modes of one or less photons. The
measured space of the $x$-setup has two possible time-bins and two
possible polarization modes, thus $H^{B_x}=H_5$ as well, however,
the two time-bins for this setup are $t_0$ and $t_1$. Following
Definition~\ref{def:HB-1} we get that that the reversed space
$H^{B^{-1}}$ contains three time-bins ($t_{-1}$, $t_0$ and $t_1$)
with two polarization modes in each, therefore $H^{B^{-1}} = H_7$,
under the single-photon assumption. The QSoP, following
Definition~\ref{def:HP} equals $H^{B^{-1}}$ since $H^A \subset
H^{B^{-1}}$.

{\bf The Attack.} Eve exploit the larger space by sending ``fake''
states using the external time bins ($t_{-1}$ and $t_1$). Eve
randomly chooses a basis, measures the qubit sent by Alice, and
sends Bob the same polarization state she found, but at $t_{-1}$
if she have used the $x$ basis, or at $t_{1}$ if she have used the
$z$ basis. Since no ancilla is kept by Eve, this is an
intercept-resend QSA.

Bob will get the same result as Eve if he uses the same basis, 
or {\em a loss} otherwise.
The mathematical description of the
attack is as follows: Eve can generate superpositions of states of
the form $\fet{{V_{t_{-1}}}{H_{t_{-1}}}
{V_{t_0}}{H_{t_0}}{V_{t_1}}{H_{t_1}}}$, where the index $\{H, V
\}$ denotes this mode has Vertical or Horizontal polarization, and
its subscript denotes the time-bin of the mode. Eve's
measure-resend attack is described as measuring Alice's qubit in the
$x$ basis, creating a new copy of the measured qubit, and
performing the transformation $(\fet{001000} \rightarrow
\fet{100000})$; $(\fet{000100} \rightarrow \fet{010000})$ or as
performing a measurement in the $z$ basis, and performing the
transformation $(\fet{001000} \rightarrow \fet{000010})$;
$(\fet{000100} \rightarrow \fet{000001})$ on the generated copy.

\vskip 12 pt
{\bf A short summery}
%
%
We see that Eve can ``force''  a desired value (or a loss) on Bob,
thus gaining all the information while inducing no errors (but
increasing the loss rate). Bob can use a shutter to block the
irrelevant time-bins but such a shutter could generate a similar
problem in the frequency domain. This attack is actually a special
case of the Trojan-pony attack, in which the imperfections of
Bob's detectors allow Eve to send states that will be un-noticed
unless the measured basis equals to Eve's chosen basis.


%
\section{Interferometric BB84 and 6-state Protocols}
\label{sec:InterferoBB84}
%
%
In order to demonstrate the power of QSA,
and to see its advantages, this section presents
a partial security analysis
of some interferometric BB84 and 6-state schemes.
Interferometric schemes are more common than any other type
of implementation in QKD
experiments~\cite{T94,MT95,GRTZ02,BBN03,DLH06,MHHTG97}
and products~\cite{idQ,magiQ}.
In this section we define the specific equipment used by Bob,
and we formulate $\calu_B$ and Bob's measurements. We then find
the spaces $H^A$, $H^{B_j}$,
$H^{B^{-1}}$ and the QSoP, $H^P$.
Finally, we demonstrate a novel attack which is found to be very
successful
against a specific variant of the BB84 interferometric scheme;
this specific QSA, which we call
the ``reversed-space attack'', is designed using the
tools
developed in Sections~\ref{sec:QSoP} and~\ref{sec:QSA}.


\subsection{Bob's equipment}\label{sec:xy-setup}

%
%
We begin with a description of interferometric
(BB84 and six-state) schemes,
which is based on sending phase-encoded qubits
arriving in two time-separated modes~\cite{T94, MT95}.
Alice encodes her qubit using two time-bins $t'_0$ and $t'_1$,
where a photon in the first mode, $\fet{10}_{t'_0t'_1}$,
represents the state $\ket{0_z}$, and a photon in the other mode,
$\fet{01}_{t'_0t'_1}$, represents $\ket{1_z}$.
The BB84 protocol of~\cite{T94, MT95} (and many others)
uses the $x$ and $y$ bases, meaning that Alice (ideally) sends
one of the following four states:
$\ket{0_x} =  (\fet{10}_{t'_0t'_1}+\fet{01}_{t'_0t'_1} ) /\sqrt{2}$;
$\ket{1_x} =  (\fet{10}_{t'_0t'_1}-\fet{01}_{t'_0t'_1} ) /\sqrt{2}$;
$\ket{0_y} =  (\fet{10}_{t'_0t'_1}+i\fet{01}_{t'_0t'_1} ) /\sqrt{2}$; and
$\ket{1_y} =  (\fet{10}_{t'_0t'_1}-i\fet{01}_{t'_0t'_1} ) /\sqrt{2}$.

\begin{figure}[p]
 \centering
 \includegraphics[width=0.8\columnwidth]{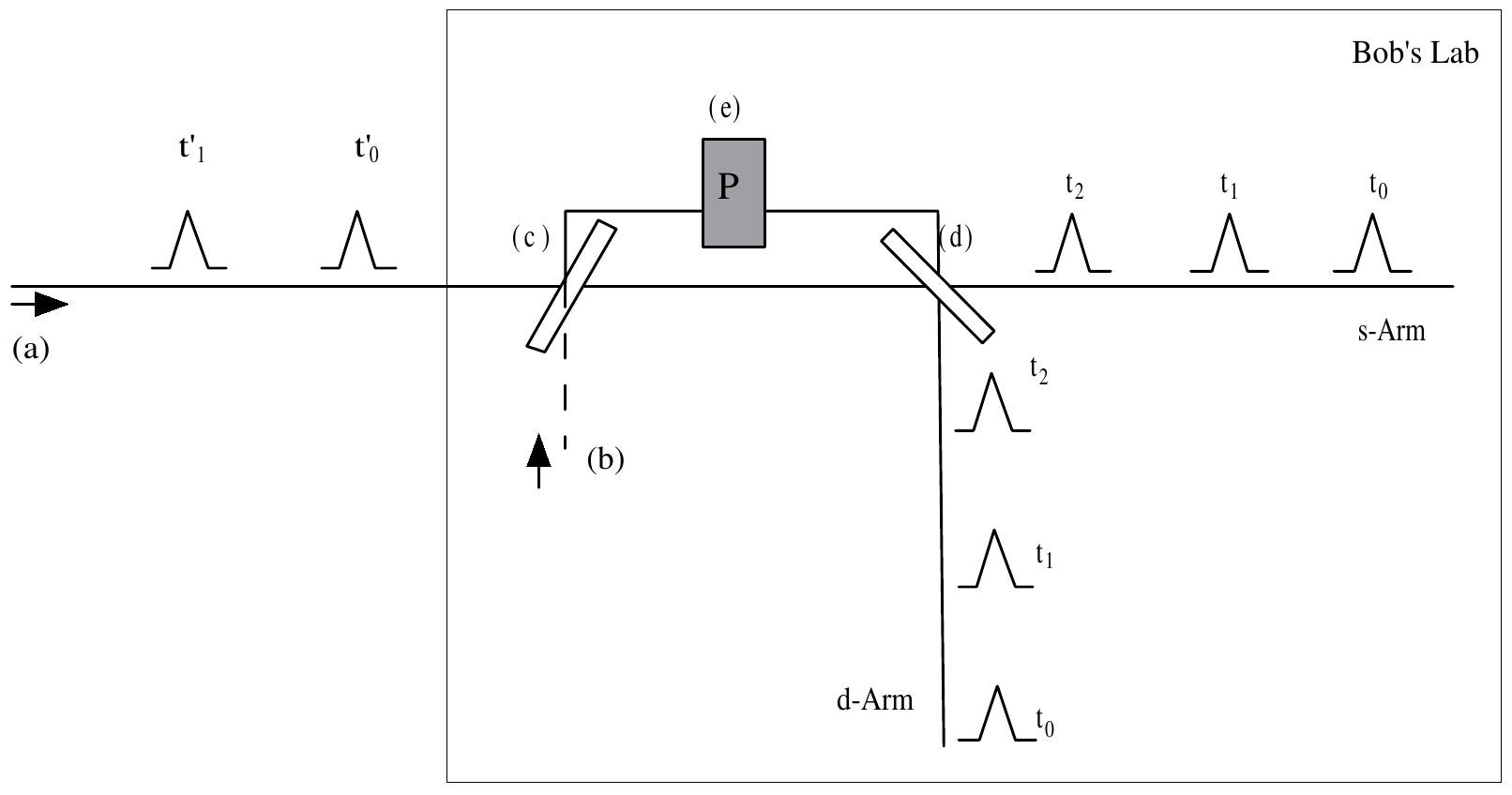}
 \caption
 {Bob's laboratory setup for the $x$ and $y$ basis.
  (a) Alice sends a qubit; (b) Vacuum states are added in the interferometer;
  (c), (d) beam-splitters; (e) phase shifter $P_\phi$.
 }
 \label{fig:lab-xy}
\end{figure}

%
%
Bob uses an interferometer built from two beam
splitters with one short path and one long path (Figure~\ref{fig:lab-xy}).
A pulse of light travels through the short arm of the interferometer
in $T_{\rm short}$ seconds, and through the long arm in $T_{\rm long}
= T_{\rm short} +\Delta T$
seconds, where $\Delta T$ is also {\em precisely} the time
separation between the two arriving modes of the qubit, $\Delta T =
t'_1 - t'_0$.
A controlled phase shifter
$P_\phi$, is placed in the long arm of the interferometer.
It performs a phase
shift by a given phase $\phi$,
i.e. $P_\phi(\ket{\psi}) = e^{i\phi}\ket{\psi}$.
The phase shifter is set to $\phi=0$ ($\phi = \pi /2 $)
when Bob measures the $x$ ($y$) basis.
%
%
Each beam splitter interferes
two input arms (modes 1, 2) into two output arms (modes 3, 4),
in the following way (for a single photon):
$\fet{10}_{1,2} \mapsto
 \frac{1}{\sqrt{2}}\fet{10}_{3,4}+\frac{i}{\sqrt{2}}\fet{01}_{3,4}$, and
$\fet{01}_{1,2} \mapsto
 \frac{i}{\sqrt{2}}\fet{10}_{3,4}+\frac{1}{\sqrt{2}}\fet{01}_{3,4}$.
The photon is
transmitted/reflected with a probability of $50\%$;
The transmitted part keeps the same phase as
the incoming photon, while the reflected
part gets an extra phase of $e^{i\pi/2}$, if it carries a single photon.
When a single mode, carrying at least a single photon,
enters a beam splitter
from one arm, and nothing enters the other input arm, we must
consider the other entry to be an additional mode (an ancilla)
in a vacuum state.

%
%
When a single mode (carrying one or more photons) enters the interferometer
at time $t'_0$,
see Figure~\ref{fig:lab-xy}, it yields
two modes at time $t_0$ due to traveling through the short arm,
and two modes at time $t_1$ due to traveling through the long arm.
Those four output modes are:
times $t_0$, $t_1$ in the `$s$' (straight) arm of the interferometer,
and times $t_0$, $t_1$ in the `$d$' (down) arm.
A basis state in this Fock space is then
$\fet{n_{s_0}, n_{s_1}, n_{d_0}, n_{d_1}}$.
In the case of having that single mode carrying exactly a single photon,
the transformation, which requires three additional empty ancillas\footnote
{See a brief description in Appendix~\ref{app:interferometer}.},
is $\fet{1}_{t'_0}\fet{000} \mapsto
(\fet{1000}-\fet{0100}+i\fet{0010} +i\fet{0001}) \thickspace / 2$.
Note that
a pulse which is sent at
a different time (say, $t'_x$)
results in the same output state, but with the appropriate
delays, i.e.\ 
\begin{equation}
\label{eqn:pulse_in_inerferometer}
\fet{1}_{t'_x}\fet{000} \mapsto
(\fet{1000}-\fet{0100}+i\fet{0010} +i\fet{0001}) \thickspace / 2\text{,}
\end{equation}
where the resulting state is defined in the Fock space
whose basis states are  $\ket{n_{s_x}, n_{s_{x+1}}, n_{d_x}, n_{d_{x+1}}}$.

%
%
Let us now examine any superposition of two modes ($t'_0$ and $t'_1$)
that enter the interferometer one after the other,
with exactly the same time difference $\Delta T$ as the
difference lengths of the arms.
The state evolves in the following way (see Appendix~\ref{app:modesEvo}):
\begin{multline}
 \cos\theta\fet{10}_{t'_0t'_1}\fet{0000} +
\sin\theta e^{i\varphi}\fet{01}_{t'_0t'_1}\fet{0000}
\mapsto
\\
\Bigl (
\cos\theta\fet{100000}_B +
 (-\cos\theta e^{i\phi} + \sin\theta e^{i\varphi})\fet{010000}_B -
 \sin\theta e^{i(\varphi+\phi)}\fet{001000}_B
\\ +
i\cos\theta\fet{000100}_B +
i(\cos\theta e^{i\phi}+\sin\theta e^{i\varphi})\fet{000010}_B +
i\sin\theta e^{i(\varphi +\phi)}\fet{000001}_B \Bigr ) /2
\label{eqn:interf_evu}
\end{multline}
describing the evolution
for any possible BB84 state sent by Alice ($\ket{0_x}$, $\ket{1_x}$,
$\ket{0_y}$, $\ket{1_y}$ determined by the value of $\varphi = 0$,
$\pi$,  $\frac{\pi}{2}$, $\frac{3\pi}{2}$ respectively,
when $\theta=\frac{\pi}{4}$).
As a result of this precise timing, these two modes are transformed into
a superposition of 6 possible modes (and not 8 modes) at the outputs,
due to interference at the second beam splitter.
Only four vacuum-states ancillas (and not six)
are required for that process.
The resulting 6 modes are $t_0$, $t_1$, $t_2$ in the `$s$' arm
and in the `$d$' arm of the interferometer.
Denote this Fock space as $H^B$, with basis elements
$\fet{n_{s_0}, n_{s_1}, n_{s_2}, n_{d_0}, n_{d_1}, n_{d_2}}_B$.

%
%
The measurement is performed as follows:
Bob opens his detectors at time $t_1$ in both
output arms of the interferometer.
A click in the ``down'' direction
means measuring the bit-value $0$, while
a click in the ``straight'' direction  means $1$.
The other modes are commonly considered as a
loss (they are not measured)
since they give an inconclusive result
regarding the original qubit.
We refer this BB84 variant as ``$xy$-BB84''.

%
%
One might want to use the $z$ basis in his QKD protocol
(using $\varphi = 0$, and $\theta=0$ or $\theta = \frac{\pi}{2}$),
for instance, in order to avoid the need for a controlled
phase shifter or for another equipment-related reason,
or in order to perform
``QKD with classical Bob''~\cite{BKM07}.
%
%
A potentially more important reason might be to perform
the 6-state QKD~\cite{Brus98, BG99,L01} protocol, due to its
improved immunity against errors (27.4\% errors versus only 20\% in
BB84~\cite{C02}).
A possible and easy to implement variant for realizing a measurement
in the $z$ basis is the following:
Bob uses the setup $\calu_{B_x}$
(i.e.\ he sets $P_\phi$ to  $\phi = 0$),
and opens his detectors at times $t_0$ and $t_2$,
corresponding to the bit-values $0$ and $1$ respectively
(See Equation~(\ref{eqn:interf_evu})).
Unfortunately, technological limitations,
e.g.\ of telecommunication
wavelength (IR) detectors,
might make it difficult for Bob
to open his detectors for more than
a single detection window per pulse.
Bob could perform a measurement of {\em just} the states
$\{\fet{000100}_B,\fet{001000}_B \}$,
opening the $d$ arm detector at time $t_0$
(to measure $\ket{0_z}$) and the $s$ arm detector
at time $t_2$ (to measure $\ket{1_z}$).
We refer this variant as ``$xyz$-six-state''.


\subsection{The quantum space of the interferometric protocols}
\label{sec:ibb84-QSOP}
%
%
We assume Alice to be almost ideal, having the realistic space $H^A=H_3$
(a qubit or a vacuum state), using two time-bin modes.
%
%
%
As we have seen,
four ancillary modes in vacuum states are added to each transmission.
Therefore, the interferometer setups $\calu_{B_x}$ and $\calu_{B_y}$
transform the 2-mode states of $H^A$ into a subspace that
resides in the 6 modes space $H^B$.
For simplicity, we assume that Eve does not generate $n$-photon states,
with $n \ge 2$,
so we can ignore high photon numbers in the $H^B$ space\footnote
{As mentioned in Section~\ref{sec:QSoP}, this assumption is not legitimate
 when proving unconditional security of a protocol.}.
Therefore, we redefine $H^B = H_7$, the space spanned by the vacuum,
and the six single-photon terms in each of the above modes.





%
%
Using the $x$ and $y$ bases,
Bob measures only time-bin $t_1$, so his actual measured spaces
consist of two modes: time-bin $t_1$ in the `$s$' arm and the `$d$' arm.
In that case, the measured spaces are  $H^{B_x} = H^{B_y} = H_3$,
spanned by the states 
$\{\fet{000000}_B$, $\fet{010000}_B$, $\fet{000010}_B \}$.
When Bob uses the $z$ basis,
he measures two different modes, so $H^{B_z}$ is spanned
by the states $\{\fet{000000}_B$, $\fet{000100}_B$, $\fet{001000}_B \}$.

%
%
Let us define the appropriate space
$H^{B^{-1}}$ for the 6-state protocol,
according to Definition~\ref{def:HB-1+anc}.
The space $H^{B^{-1}}$ is spanned by the states given by performing
$\calu \in \{ \calu_{B_x}, \calu_{B_y} \}$
on  $\{\fet{000000}_{B}$, $\fet{010000}_{B}$, $\fet{000010}_{B} \}$,
as well as the states given by performing $\calu_{B_z}$ on
$\{\fet{000000}_{B}$, $\fet{000100}_{B}$ , $\fet{001000}_{B} \}$.
Interestingly, once applying $\calu^{-1}$, the resulting states
are embedded in an 8-mode space
defined by the two
incoming arms of the interferometer, `$a$' (from Alice) and `$b$' (from Bob),
at time bins $t'_{-1}$, $t'_{0}$, $t'_{1}$, and $t'_{2}$. The basis states
of $H^{B^{-1}}$ are listed in Appendix~\ref{app:interf_HB-1_basis}.
Following Definition~\ref{def:HP+anc},
the QSoP $H^P$ of this implementation for the 6-state protocol,
is the subsystem of $H^{B^{-1}}$
which is {\em controlled} by Eve.
It is spanned by the 8-mode states spanning
$H^{B^{-1}}$ after tracing out Bob.
The space that contains those ``traced-out'' states
has only four modes that are controlled by Eve, specifically, input `$a$'
of the interferometer at times $t'_{-1}$ to $t'_2$,
having a basis state of the form
$\fet{a_{t'_{-1}}a_{t'_0}a_{t'_1}a_{t'_2}}_P$.
Given the single-photon restriction, we get
$H^P = H_5$, namely,
the space spanned by the vacuum state,
and a single photon in each of the four modes, i.e.\
 $\{ \fet{0000}_{P}$, $\fet{1000}_{P}$,
 $\fet{0100}_P$, $\fet{0010}_P$ , $\fet{0001}_P \}$.
This same result is obtained also if Bob measures all the
six modes in $H^B$. 

%
%
Bob might want to see how the basis states of the
4-mode QSoP, $H^P$,
evolve through the interferometer 
in order to place detectors on the resulting modes, which
will be used to identify Eve's attack.
It is interesting to note, that those basis states result in
{\em 10 different non-empty modes (!)}.
If Bob measures all these modes, he {\em increases} the QSoP,
and maybe allows Eve to attack a larger space,
and so on and so forth.
Therefore, in order to perform a security analysis,
one must first fix the scheme and only then assess the QSoP.
Otherwise, a ``ping-pong'' effect might
increase the spaces' dimensions to infinity.
%
%
A similar, yet reversed logic, hints that it could actually be better
for Bob, in terms of the simplicity of the analysis for
the ``$xy$-BB84'' scheme, to measure
{\em just} the two modes at $t_1$
(i.e.\ the space spanned by $\fet{0,n_{s_1},0,0,n_{d_1},0}_B$),
thus reducing the QSoP  to a 2-mode space, $H^P=H^A$, see 
Appendix~\ref{app:QSoP2ModesIntf}.
Although Eve is allowed to attack a larger
space than this two-mode $H^P$, she has no advantage in doing so:
pulses that enter the interferometer on different modes
(i.e.\ other time-bins than $t'_0$ and $t'_1$), never interfere
with the output pulses of time-bin $t_1$ measured by Bob. Therefore,
state occupying different modes can not
be distinguished from the states in which those modes are empty.

\subsection{The ``Reversed-Space'' attack on interferometric protocols}
\label{sec:RSA}


%
%
Consider a BB84 variant 
in which Bob uses only the $x$ and the $z$ bases,
using a single interferometer, where the $z$-basis measurement
is performed according to the description in the last few lines
of Section~\ref{sec:xy-setup}. We refer this variant as
``$xz$-BB84''. The QSoP of this scheme, $H^P$ is the space
described above for the ``$xyz$-six-state'' protocol.
The following attack
\begin{align}
\ket{0}_{E}\ket{0100}_{P}  \mapm{\calu_{E}} &
 {\frac{1}{2}}\ket{E_0}_{ E}\bigl (\fet{1000}_{P} + \fet{0100}_{P} \bigr )
  {}+ {\frac{1}{2}}\ket{E_1}_{ E}\bigl (\fet{0010}_{P} + \fet{0001}_{P} \bigr )
\\
\ket{0}_{E}\ket{0010}_{P} \mapm{\calu_{E}} &
 {\frac{1}{2}}\ket{E_1}_{ E}\bigl (-\fet{1000}_{P} + \fet{0100}_{P} \bigr )
  {}+{\frac{1}{2}}\ket{E_0}_{ E}\bigl (\fet{0010}_{P} - \fet{0001}_{P} \bigr )
\end{align}
which we call ``the Reversed-Space Attack'', allows Eve
to acquire information about the 
transmitted qubits, without inducing {\em any} errors.
%
%
The states $\ket{\cdot}_E$ denote Eve's ancilla
which is not necessarily a photonic system.
The state $\ket{0_z}_{A} \equiv \fet{0100}_P$ and
$\ket{1_z}_{A} \equiv \fet{0010}_P$ are the regular states
send by Alice, where we added the relevant extension
of $H^A$ in $H^P$.
When $\ket{0_z}_{A}$ is sent by Alice,
the attacked state $\calu_{E} \ket{0}_{E}\ket{0_z}_{A}$ 
reaches Bob's interferometer, and interferes
in a way such that it can never reach Bob's detector at time $t_2$,
i.e. $\fra{001000}_B \calu_{B_x} \left(
\left (\calu_{E}\ket{0}_{E}\ket{0_z}_{A} \right )\fet{0000}_{B'} \right )=0$.
Although the attacked state $\calu_{E}\ket{0}_{E}\ket{0_z}_{A} $
reaches modes that Alice's original state $\ket{0_z}_{A}$ can never
reach, Bob never measures those modes, and cannot notice the attack.
A similar argument applies when Alice sends $\ket{1_z}_{A}$.

%
%
As for the $x$ basis\footnote{For simplicity 
we use the shorter notation 
$\ket{0_x}\equiv(\fet{0100}_P+\fet{0010})/\sqrt{2}$, etc.},
this attack satisfies
\begin{multline}
\ket{0}_{E}\ket{0_x}_{A} \mapsto
\\
\frac{1}{\sqrt{8}}
(\ket{E_0}_{ E} + \ket{E_1}_{ E})
(\fet{0100}_P + \fet{0010}_P) +
\frac{1}{\sqrt{8}}
(\ket{E_0}_{ E} - \ket{E_1}_{ E})
(\fet{1000}_P - \fet{0001}_P) \phantom{1\text{.}}
\end{multline}
\begin{multline}
\ket{0}_{E}\ket{1_x}_{A} \mapsto
\\
\frac{1}{\sqrt{8}}
(\ket{E_0}_{ E} - \ket{E_1}_{ E})
(\fet{0100}_P -  \fet{0010}_P) +
\frac{1}{\sqrt{8}}
(\ket{E_0}_{ E} + \ket{E_1}_{ E})
(\fet{1000}_P + \fet{0001}_P) \text{.}
\end{multline}
The first element in the sum results in the desired interference in Bob's lab,
while the second is not measured by Bob's detectors at time $t_1$.
By letting Eve's probes $\ket{E_0}_{E}$ and $\ket{E_1}_{E}$ be orthogonal states,
Eve gets a lot of information while inducing no errors at all.
%
Yet, we find that Eve is increasing the loss rate by this attack to
87.5\%, but a very high loss rate is anyhow expected by Bob
(as explained in the analysis of the PNS~\cite{BLMS00}
 and the tagging~\cite{GLLP04} attacks).

%
%
In conclusion,
this attack demonstrates the risk of using various
setups without giving full security analysis for the {\em specific} setup.
We are not familiar
with any other security analysis that takes into account the enlarged
space generated by the inverse-transformation of
Bob's space.

\section{Conclusion}
\label{sec:conclusion}
%
%
In this paper we have defined the QSA,
a novel attack that generalizes all currently
known attacks on the channel.
This new attack brings a new
method for performing security analysis of protocols.
The attack is based on a realistic view of
the quantum spaces involved, and in particular, the spaces
that become larger than the theoretical ones,
due to practical considerations.
Although this paper is explicitly
focused on the case of uni-directional
implementations of a few schemes, its main observations
and methods apply to any uni-directional
QKD protocol, to bi-directional QKD protocols,
and maybe also to any realistic
quantum cryptography scheme beyond QKD.

The main conclusion of this research is that the quantum space which is
attacked by Eve can be assessed, given a proper understanding of the
experimental limitations. This assessment requires a novel cryptanalysis
formalism --- analyzing the states generated in Alice's lab, as well as
the states that are to be measured by
Bob (assessing them as if they go backwards in time from Bob's lab);
this type of analysis resembles the two-time formalism in quantum
theory~\cite{AAD85,VAA87}.

%
%
Open problems for further theoretical research include:
1.-- Generalization of the QSA to other conventional
protocols (such as the two-state
protocol, EPR-based protocols, d-level protocols, etc.); such a generalization
should be rather straightforward.
2.-- Proving unconditional security (or more limited security results such as
``robustness''~\cite{BKM07}) against various QSAs. This is especially important for the
interferometric setup, where the QSoP is much larger than Alice's
six-dimensional space (the one spanned by $\chi^6$).
3.-- Describing the QSA for more complex protocols, such
as two-way protocols \cite{MHHTG97,BF02, BKM07}
in which the quantum communication is bi-directional,
and protocols which use a larger set of
states such as data-rejected protocols \cite{BHP93}
or decoy-state protocols \cite{H03,W05,LMC05, YSS07}.
4.-- Extend the analysis and results to composable QKD~\cite{BHLMO05}.
5(a).-- In some cases, if Bob uses ``counters''
and treats various measurement outcomes as errors,
the effective QSoP relevant for proving security is potentially
{\em much smaller} than the QSoP defined here.
5(b).-- Adding counters on more modes
increases the QSoP defined here, but might
allow analysis of a smaller ``attack's QSoP'',
if those counters are used
to identify Eve's attack.
More generally,
the connection between the way Bob interprets
his measured outcomes, and the
``attack's QSoP'' is yet to be further analyzed.
\paragraph{{\bf Acknowledgments.}}
We thank Michel Boyer, Dan Kenigsberg and Hoi-Kwong Lo
for helpful remarks.


\newpage

\begin{appendix}
\section*{Appendix}


\section{Mathematical Description of the PNS attack}
\label{App:MathPNS}

The PNS attack can be realized using (an infinite set of)
polarization independent beams splitters.
Eve uses a beam splitter to split photons from Alice's state.
Using a non-demolition measurement Eve measures the number of photons
in one output of the beam splitter, and repeat the splitting
until she acquires exactly one photon.
Formally $\calu_E$ is defined:
\begin{align*}
\fet{00}_E\fet{02}_A &\mapsto \fet{01}_E\fet{01}_P  &
\fet{00}_E\fet{10}_A &\mapsto \fet{10}_E\fet{00}_P \\
\fet{00}_E\fet{20}_A &\mapsto \fet{10}_E\fet{10}_P  &
\fet{00}_E\fet{01}_A &\mapsto \fet{01}_E\fet{00}_P \\
\fet{00}_E\fet{11}_A &\mapsto  (\fet{01}_E\fet{10}_P +
\fet{10}_E\fet{01}_A) / \sqrt{2}\text{.}
\end{align*}
Whenever Alice sends a pulse with two photons of
the same polarization, Eve and Bob end  up, each, with
having a single photon of the original polarization.
\begin{proposition}
Eve's PNS attack for a pulse of 2 photons,
gives Eve full information while inducing no errors.
\end{proposition}
\begin{proof}
According to its definition
it is trivial to verify the
attack for the horizontal and
vertical polarizations $\ket{0_z}^{(2)}$ and $\ket{1_z}^{(2)}$
(where $\ket{P}^{(k)}$ means $k$ photons having polarization $P$).
Using the standard creation and annihilation operators ($a^\dagger$ and $a$)\footnote{
See any quantum optics book, e.g.\ \cite{SZ97}},
we can write the state of two photons in the diagonal polarization ($x$ basis):
 $\ket{0_x}^{(2)} =
\left (\frac{1}{\sqrt{2}} (a^\dagger_1 + a^\dagger_2) \right )^2 \fet{00} =
\frac{1}{2}\bigl (\fet{20} + \sqrt{2}\fet{11} + \fet{02}\bigr ) $,
similarly $\ket{1_x}^{(2)} = \frac{1}{2}\bigl
(\fet{20} - \sqrt{2}\fet{11} + \fet{02}\bigr ) $.

\begin{eqnarray*}
\fet{00}_E\ket{0_x}^{(2)}_{ P} &\equiv & \frac{1}{2}
\fet{00}_E \bigl (\fet{20} +
\sqrt{2}\fet{11} + \fet{02}\bigr )_P \\
&\mapm{\calu_E} & \frac{1}{2} \bigl (\fet{10}_E\fet{10}_P  +
        \fet{01}_E\fet{10}_P + \fet{10}_E\fet{01}_P +
        \fet{01}_E\fet{01}_P  \bigr ) \\
& = &   \frac{1}{2}  \bigl ( (\fet{10}_E+\fet{01}_E )\fet{10}_P
 + (\fet{10}_E+\fet{01}_E )\fet{01}_P    \bigr ) \\
& = & \frac{1}{2} (\fet{10}_E+\fet{01}_E)(\fet{10}_P+\fet{01}_P) \\
 &\equiv & \ket{0_x}_{ E}\ket{0_x}^{(1)}_{ P}
\end{eqnarray*}
\begin{eqnarray*}
\fet{00}_E\ket{1_x}^{(2)}_{ P} &\equiv &
\frac{1}{2} \fet{00}_E \bigl(\fet{20} - \sqrt{2}\fet{11} + \fet{02}\bigr )_P \\
&\mapm{\calu_E}& \frac{1}{2} \bigl (\fet{10}_E\fet{10}_P  -
        \fet{01}_E\fet{10}_P - \fet{10}_E\fet{01}_P +
        \fet{01}_E\fet{01}_P  \bigr ) \\
& = &   \frac{1}{2}  \bigl ( (\fet{10}_E-\fet{01}_E )\fet{10}_P
 - (\fet{10}_E-\fet{01}_E )\fet{01}_P    \bigr ) \\
& = & \frac{1}{2} (\fet{10}_E-\fet{01}_E)(\fet{10}_P-\fet{01}_P) \\
 &\equiv & \ket{1_x}_{ E}\ket{1_x}^{(1)}_{ P}
\end{eqnarray*}
Which completes the proof.
\end{proof}

\subsection{Polarization change}
\label{app:polU}
A Polarization based QKD protocol makes a use of a Pockel cell ($\calu_{B_x}$),
rotating the polarization of the photons going through it.
For a single photon, its action is trivial,
\begin{align}
\nonumber
\fet{10} &\mapm{\calu_{B_x}} 
  \frac{1}{\sqrt{2}} \left (\fet{10}+\fet{01} \right )\text{, and} \\
\nonumber
\fet{01} &\mapm{\calu_{B_x}} 
  \frac{1}{\sqrt{2}} \left (\fet{10}-\fet{01} \right )\text{.} \\
\end{align}
For a state that contains multiple photons, the transformation is not
intuitive, and most simply defined using the creation and annihilation operators.
In a somewhat simplified way, the Pokcel cell
can be considered as performing
$a^\dagger_1 \mapsto \left (\frac{1}{\sqrt{2}} (a^\dagger_1 + a^\dagger_2) \right )$
and
$a^\dagger_2 \mapsto \left (\frac{1}{\sqrt{2}} (a^\dagger_1 - a^\dagger_2) \right )$,
so that a state is transformed in the following way
\begin{align}
\fet{nm} = {\left (a^\dagger_1 \right)}^n {\left (a^\dagger_2 \right )}^m
\fet{00} \mapm{\calu_{B_x}}
{\left (\frac{1}{\sqrt{2}} (a^\dagger_1 + a^\dagger_2) \right )}^n
{\left (\frac{1}{\sqrt{2}} (a^\dagger_1 - a^\dagger_2) \right )}^m\fet{00}
\text{.}
\end{align}


\section {QSoP of the Interferometeric Scheme: Supplementary Information}

\subsection{A (brief) graphical description of pulses evolution through interferometer}
\label{app:interferometer}
\begin{figure}[!hp]
     \centering
     \subfigure[Time $T_0$: the pulse (1) is about to enter the interferometer.
        A vacuum ancilla (2) is added in the input of the first beam splitter.]{
          \label{fig:1evo-t0}
          \includegraphics[width=.45\textwidth]{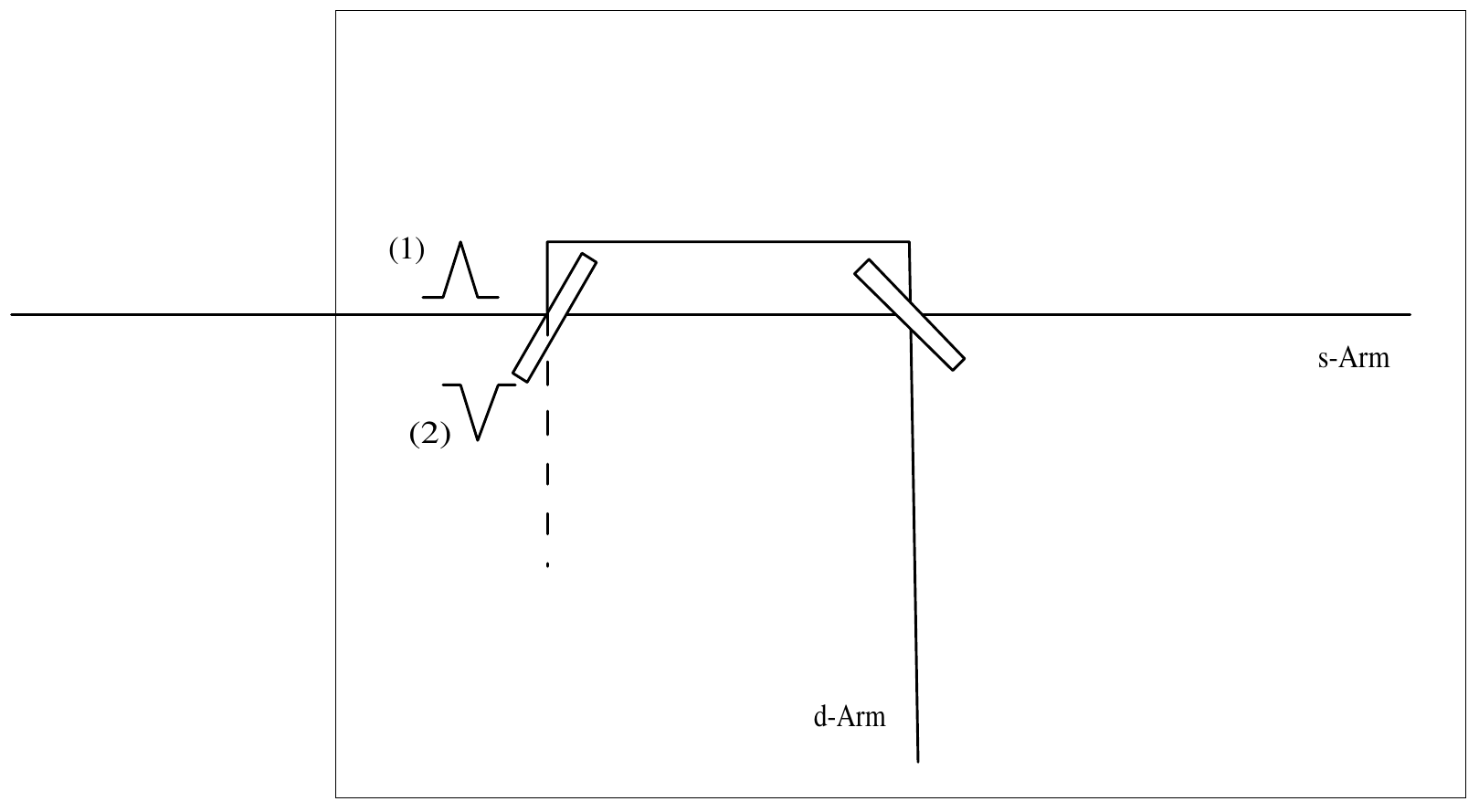}}
     \hspace{.3in}
     \subfigure[Time $T_1$: Pulses (1) and (2) interfere and become
a superposition of (3) and (3') in
the short and long arms of the interferometer, respectively,
$\ket{1}_1\ket{0}_2 \maps{BS}
(\ket{1}_3\ket{0}_{3'} + i\ket{0}_3\ket{1}_{3'})/\sqrt{2}$.
Pulse (3) is about to enter the second beam splitter so a vacuum ancilla  is added (4).]{
          \label{fig:1evo-t1}
          \includegraphics[width=.45\textwidth]{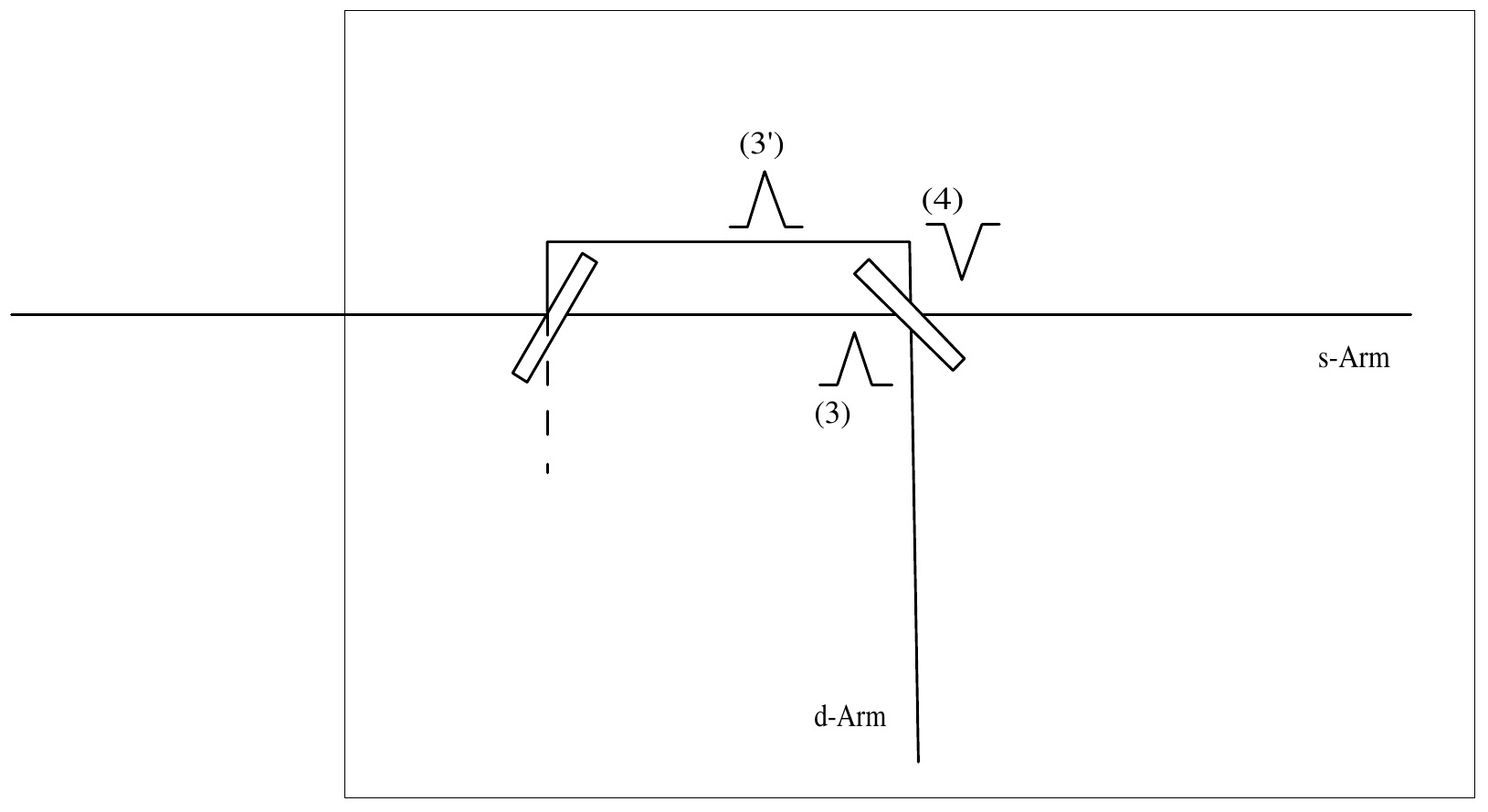}} \\
     \vspace{.3in}
     \subfigure[Time $T_2$: pulses (5) and (5') are created by pulses (3) and (4),
$\frac{1}{\sqrt{2}}\ket{0}_4\ket{1}_3 \maps{BS}
(i\ket{1}_5\ket{0}_{5'}+\ket{0}_5\ket{1}_{5'})/2$.
 Pulse (3') is about to enter the second beam-splitter so a vacuum ancilla is added (6).]{
           \label{fig:1evo-t2}
           \includegraphics[width=.45\textwidth]{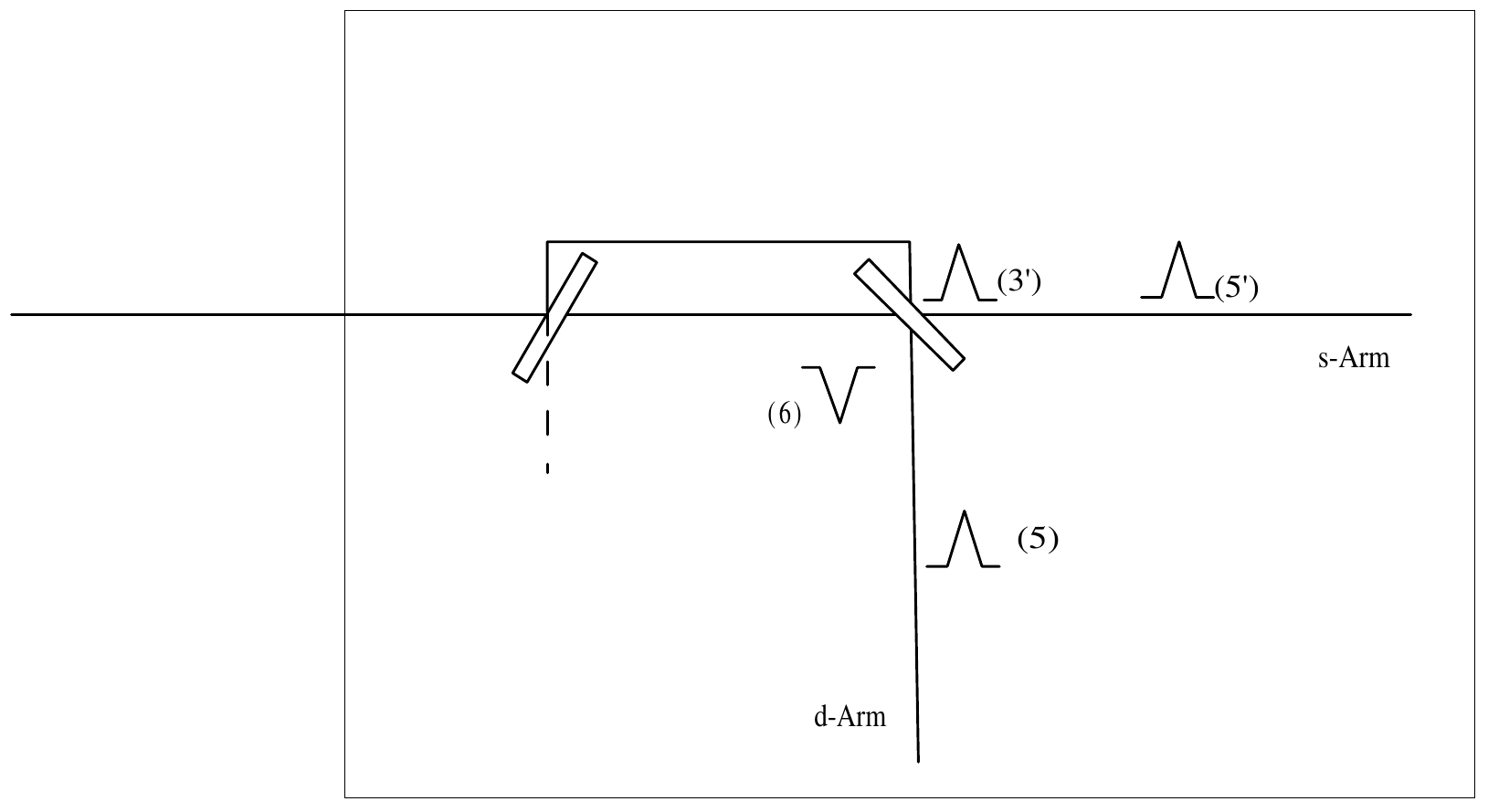}}
     \hspace{.3in}
     \subfigure[Time $T_3$: Pulses (7) and (7') are
created by interfering (3') and (6).  $\frac{i}{\sqrt{2}}\ket{1}_{3'}\ket{0}_6
 \maps{BS} (i\ket{1}_7\ket{0}_{7'} - \ket{0}_7\ket{1}_{7'})/2$.]{
          \label{fig:1evo-t3}
          \includegraphics[width=.45\textwidth]{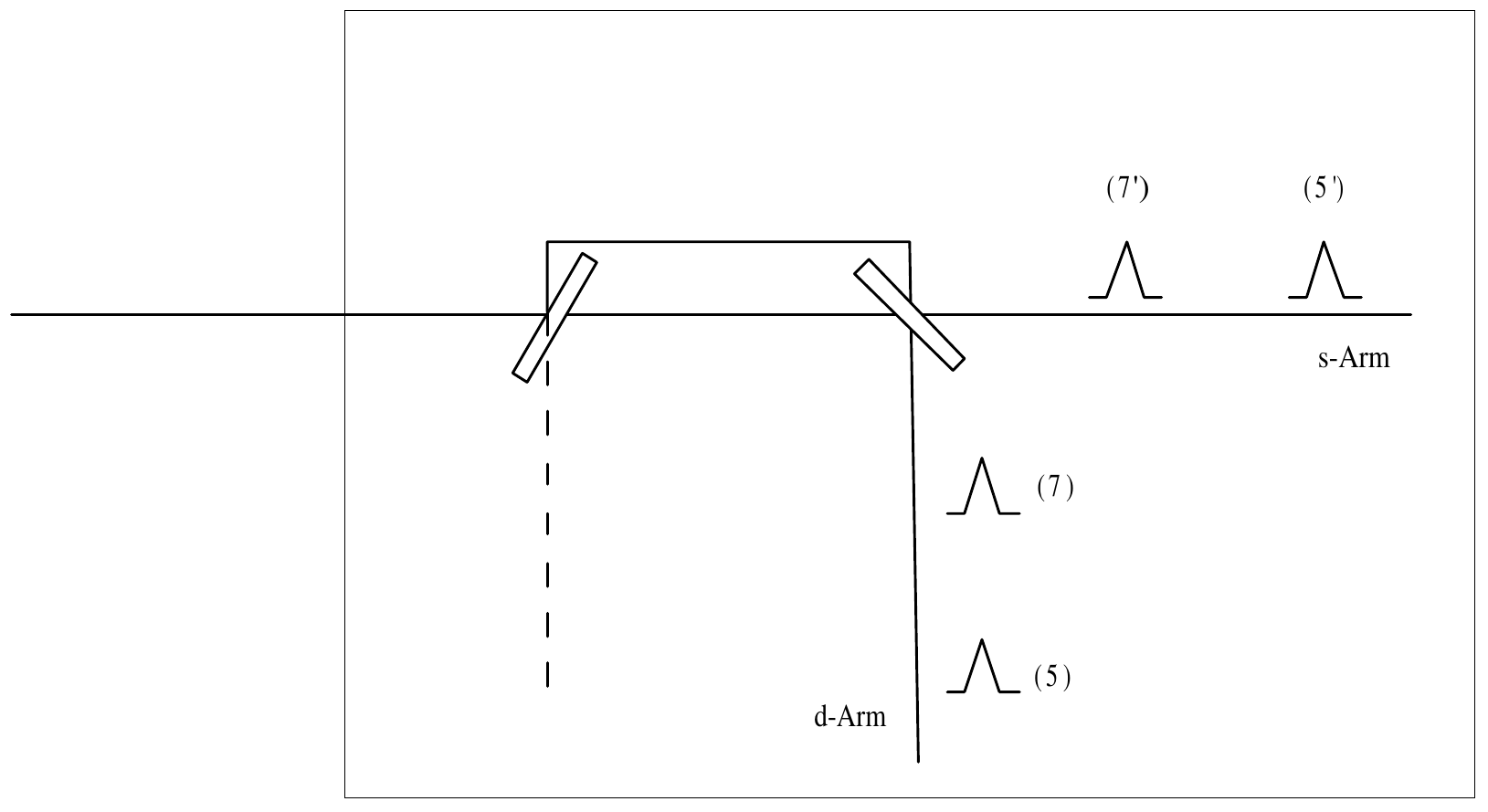}}
     \caption{Evolution in time of a single photon pulse through an interferometer
     satisfying $\ket{1000}_{1,2,4,6} \maps{Interferometer}
(\ket{1000}_{5',7',5,7} - \ket{0100}_{5',7',5,7}+
i\ket{0010}_{5',7',5,7} + i\ket{0001}_{5',7',5,7} )/2$.
The numbers represent the appropriate mode number of each pulse.
The input state ($\ket{1}_{t_0}\ket{000}$) consists of modes
(1) for the pulse at $t_0$ and (2), (4) and (6) for the vacuum ancillas.
The output modes that correspond to the state
$\ket{n_{s_0}, n_{s_1}, n_{d_0}, n_{d_1}}$
are modes (5'), (7'), (5) and (7) respectively.}
     \label{fig:1evo}
\end{figure}

\begin{figure}[!hp]
\centering
\subfigure[Time $T_0$: The general single-photon qubit ($\alpha\ket{0}+\beta\ket{1}$)
is sent to Bob is in two modes (1) and (2).
    Bob adds two vacuum ancillas (1') and (2') that
    interfere with the photon in the first beam splitter (BS-1).]{
    \label{fig:2evo-0}
    \includegraphics[width=0.45\columnwidth]{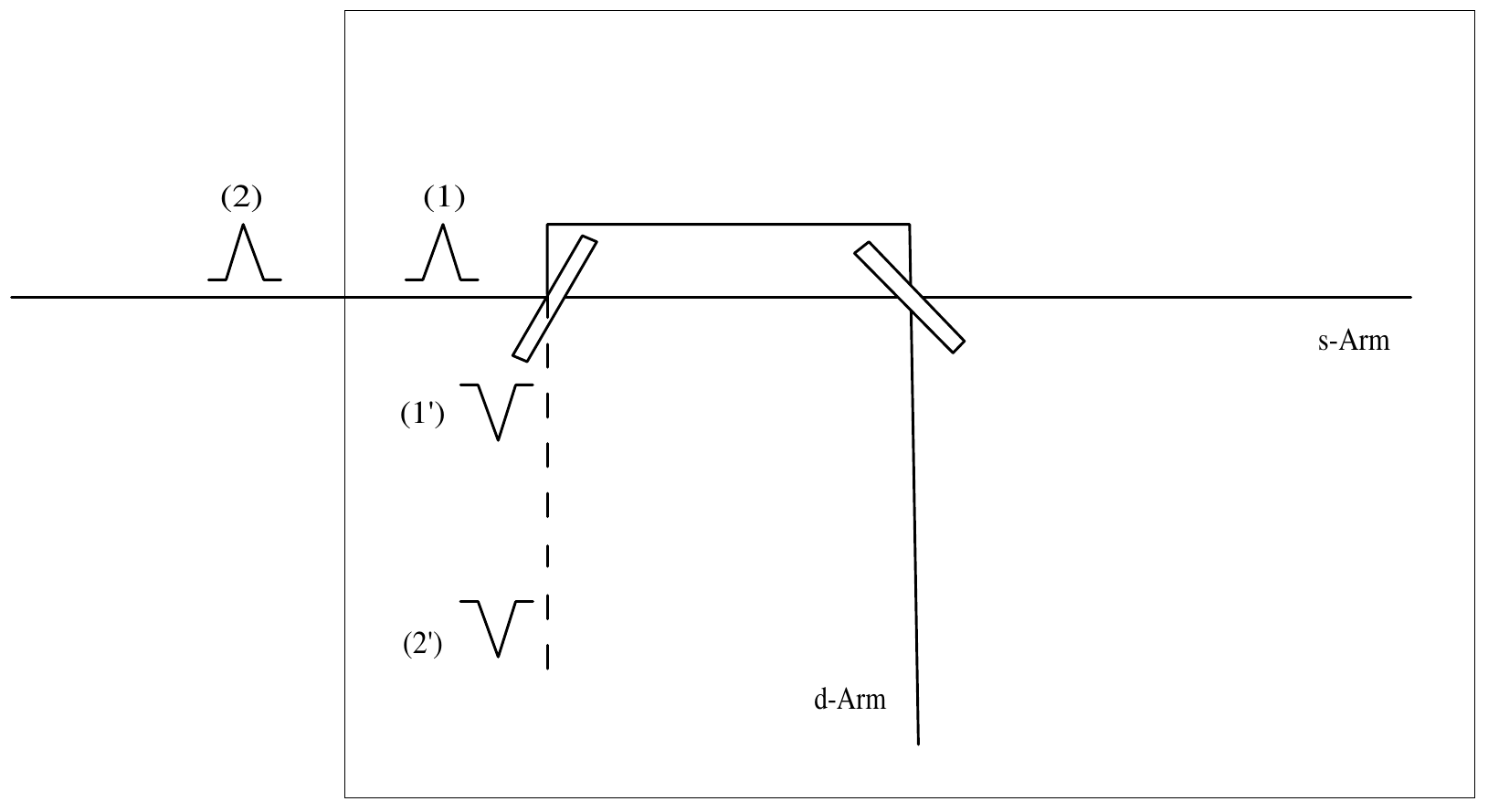}}
\hspace{0.3in}
\subfigure[Time $T_1$: Modes (1) and (1') interfere and create (3) and (3')
 in the short and long arm respectively,
$\alpha\ket{1}_1\ket{0}_{1'} \maps{BS}
\frac{\alpha}{\sqrt{2}} (\ket{1}_3\ket{0}_{3'} + i\ket{0}_3\ket{1}_{3'})$.
Pulse (3) is about to enter BS-2
so a vacuum ancilla is added (4).]{
    \label{fig:2evo-1}
    \includegraphics[width=0.45\columnwidth]{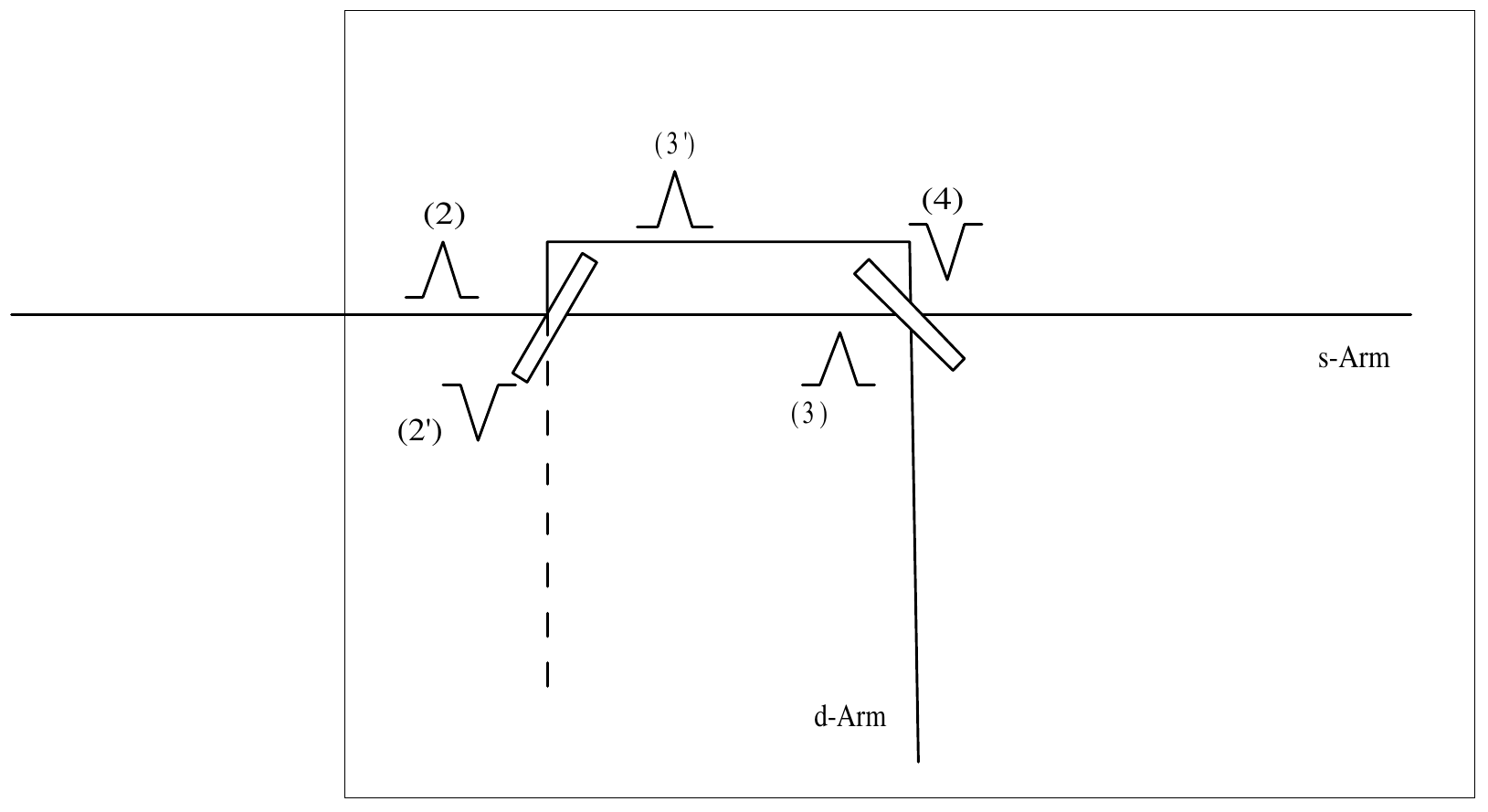}} \\
\subfigure[Time $T_2$: Pulses (5) and (5') are created by the interference of
 (3) and (4)
$\frac{\alpha}{\sqrt{2}}\ket{0}_4\ket{1}_3
 \maps{BS}
    \frac{i\alpha}{2} \ket{1}_5\ket{0}_{5'} +\frac{\alpha}{2}\ket{0}_5\ket{1}_{5'}$.
Pulses (6) and (6') created by the interference of (2) and (2') in BS-1
$\beta\ket{1}_2\ket{0}_{2'} \maps{BS}
\frac{\beta}{\sqrt{2}} (\ket{1}_6\ket{0}_{6'} + i\ket{0}_6\ket{1}_{6'})$.
.]{
    \label{fig:2evo-2}
    \includegraphics[width=0.45\columnwidth]{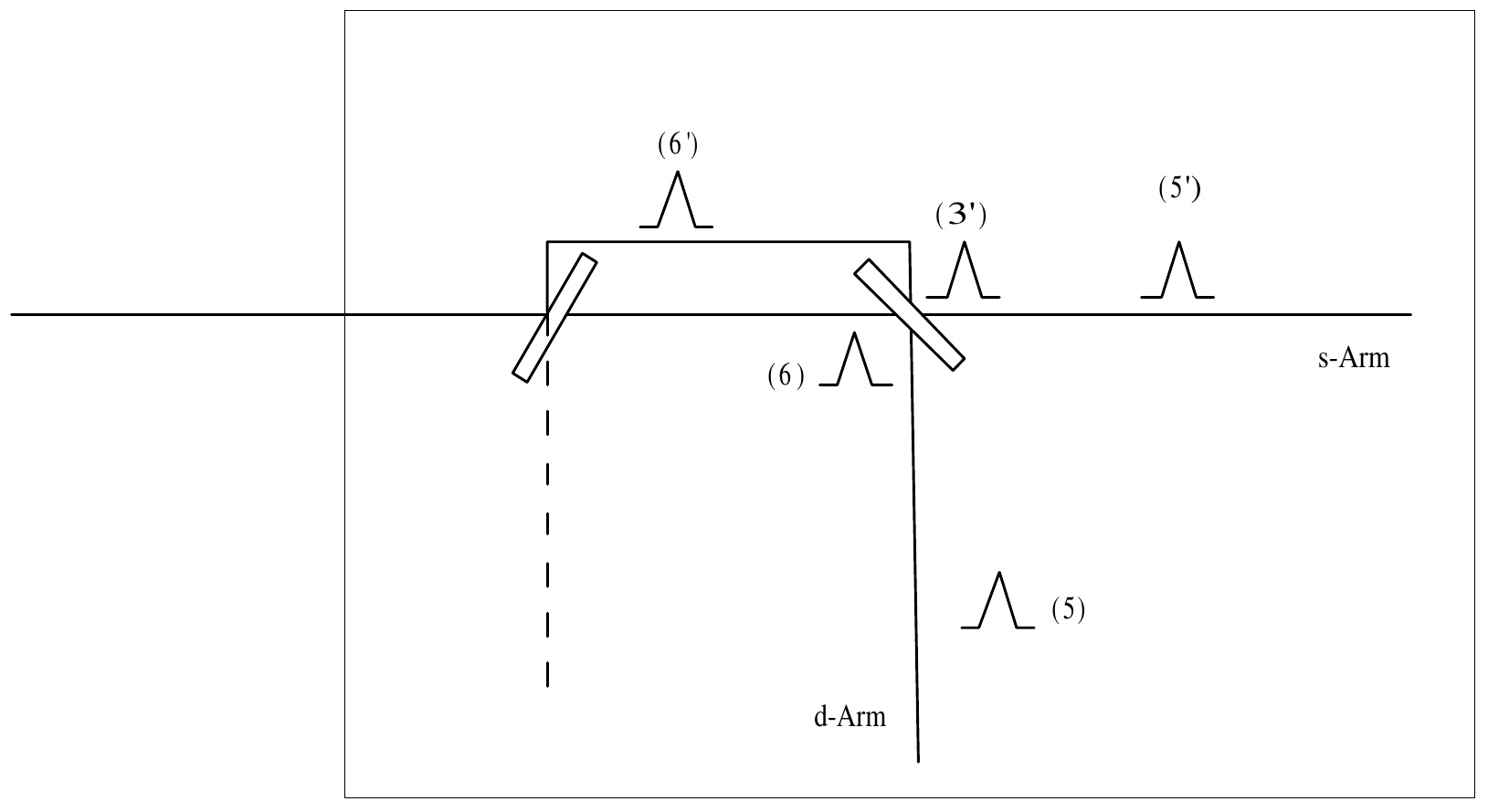}}
\hspace{0.3in}
\subfigure[Time $T_3$: Pulses (7) and (7') are created by the interference of (3')
and (6) in BS-2
$ \frac{i\alpha}{\sqrt{2}}\ket{1}_{3'}\ket{0}_6
+ \frac{\beta}{\sqrt{2}}\ket{0}_{3'}\ket{1}_6
\maps{BS}
  \frac{i(\alpha+\beta)}{2}\ket{1}_7\ket{0}_{7'}
+ \frac{\beta-\alpha}{2} \ket{0}_7\ket{1}_{7'}$ .
Pulse (6') is about to enter BS-2
 so a vacuum ancilla is added (8).]{
    \label{fig:2evo-3}
    \includegraphics[width=0.45\columnwidth]{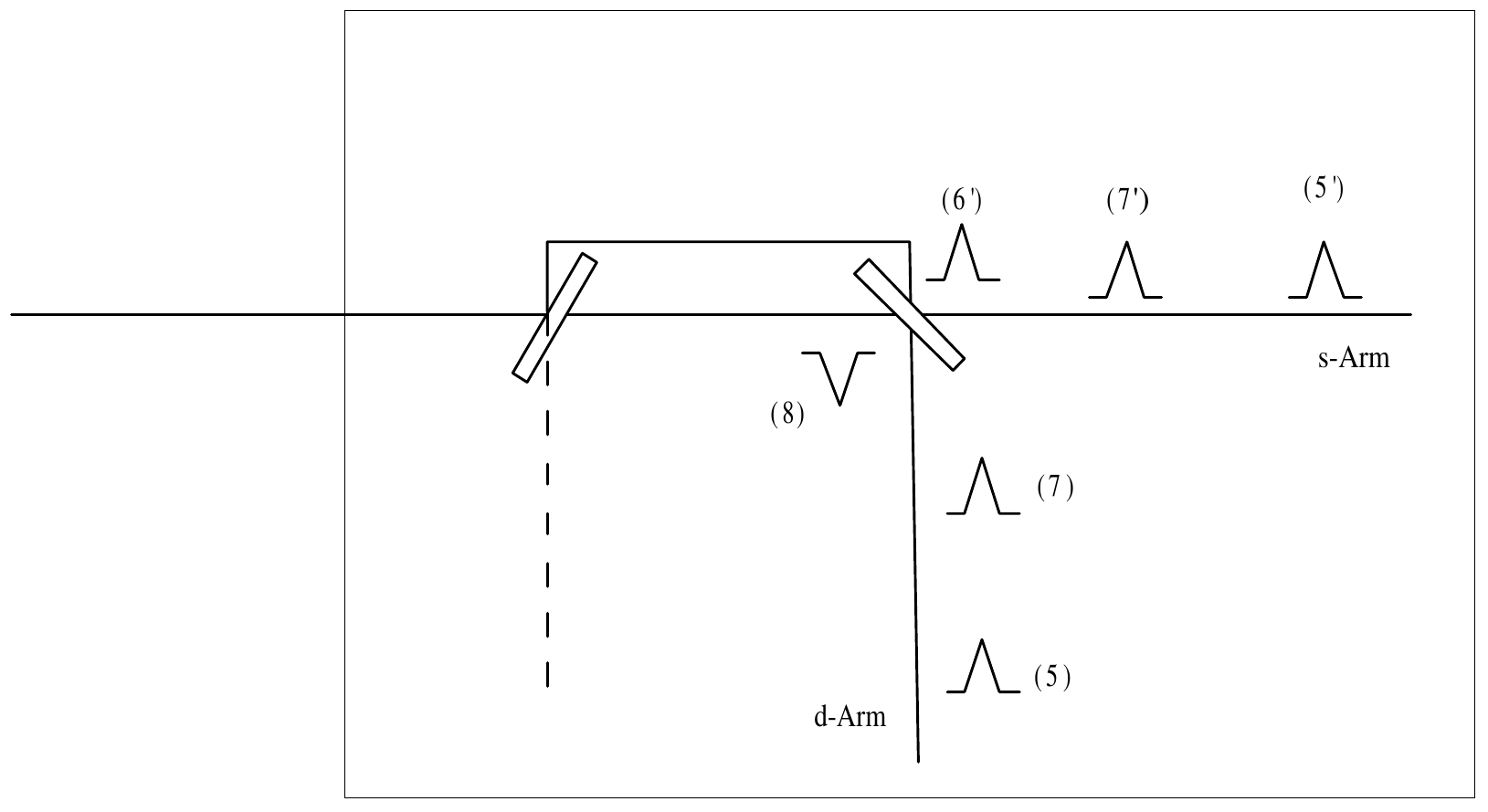}} \\
\hspace{0.1in}
\subfigure[Time $T_4$: Pulses (9) and (9') are created by the interference of (6')
and (8) in BS-2
$\frac{i\beta}{\sqrt{2}}\ket{1}_{6'}\ket{0}_{8} \maps{BS}
    \frac{i\beta}{2} \ket{1}_9\ket{0}_{9'} - \frac{\beta}{2}\ket{0}_9\ket{1}_{9'}$.
]{
    \label{fig:2evo-4}
    \includegraphics[width=0.45\columnwidth]{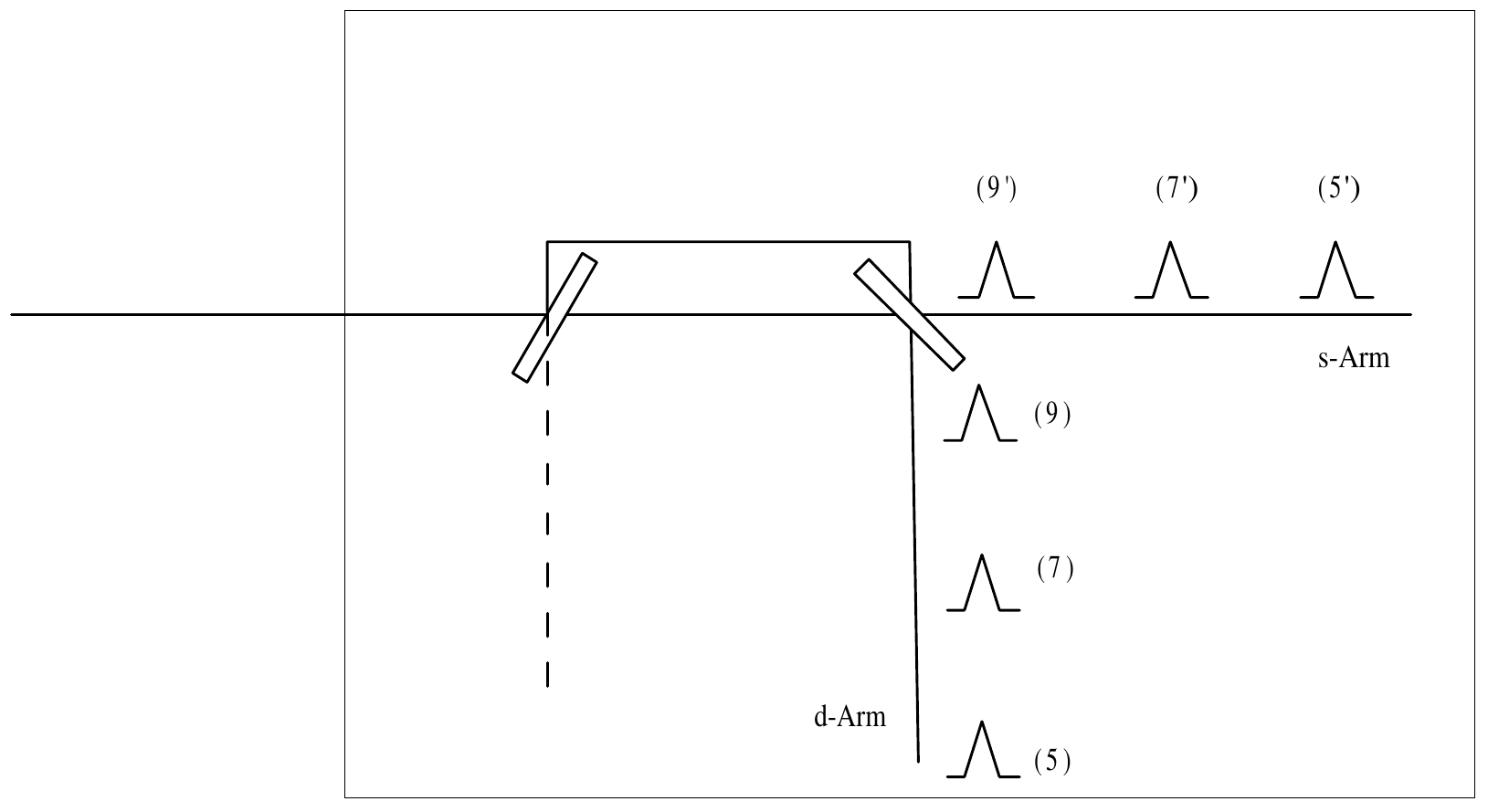}} \\
\caption{Evolution in time of two modes through an interferometer satisfying
$ (\alpha\ket{1}_1\ket{0}_2 + \beta\ket{0}_1\ket{1}_2)\ket{0000}_{1',2',4,8}
 \maps{Interferometer}
(\frac{\alpha}{2}\ket{100000}
+\frac{\beta-\alpha}{2} \ket{010000}
-\frac{\beta}{2}\ket{001000}
+ \frac{i\alpha}{2}\ket{000100}
+\frac{i(\alpha+\beta)}{2}\ket{000010}
+ \frac{i\beta}{2} \ket{000001})_{5', 7', 9', 5, 7, 9}$.
The numbers represent the appropriate mode number of each pulse. The corresponding
state is $\ket{n_{s_0}n_{s_1}n_{s_2}n_{d_0}n_{d_1}n_{d_2}}$.
}
\label{fig:2evo}
\end{figure}

See Figure~\ref{fig:1evo} for evolution of a single occupied mode through
the interferometer, and Figure~\ref{fig:2evo} for evolution of
two superpositioned modes.



\subsection{Evolution of modes through the interferometer}
\label{app:modesEvo}
In order to simplify the analysis
(a simplification that is not allowed when proving the full security of
a scheme) we look at the
ideal case in which exactly one photon (or none)
is sent by Alice.
The basis states are then the vacuum $\fet{000000}_B \equiv \fet{V}_B$,
and the six states (that we denote for simplicity by)
$\fet{100000}_B \equiv \fet{s_0}_B$;
$\fet{010000}_B \equiv \fet{s_1}_B$;
$\fet{001000}_B \equiv \fet{s_2}_B$;
$\fet{000100}_B \equiv \fet{d_0}_B$;
$\fet{000010}_B \equiv \fet{d_1}_B$ and
$\fet{000001}_B \equiv \fet{d_2}_B$.

%
%
The full transformation of a single photon pulse through the 
interferometer is given by Equation~\eqref{eqn:pulse_in_inerferometer}.
Alice sends photons at time bins $t'_0$ and $t'_1$ only, so 
the interferometer transformation on Alice's basis states is 
$\fet{00}_A\fet{0000}_{\hat B} \mapsto \fet{V}_{B}$, and
\begin{eqnarray}\label{eqnB+-On01}
\begin{array}{l}
\fet{10}_A\fet{0000}_{\hat B} \mapsto
(\fet{s_0}_{B}-e^{i\phi}\fet{s_1}_{B}+i\fet{d_0}_{B}
    +ie^{i\phi}\fet{d_1}_{B}) \thickspace / 2 \\
\fet{01}_A\ket{0000}_{\hat B} \mapsto
(\fet{s_1}_{B}-e^{i\phi}\fet{s_2}_{B}+i\fet{d_1}_{B}
    +ie^{i\phi}\fet{d_2}_{B})\thickspace / 2
\ ,
\end{array}
\end{eqnarray}
where $\ket{0000}_{\hat B}$ denotes ancilla added
during the process\footnote{Those ancillas (the space $H^{\hat B}$) are originated
by Alice extended space $H^P$ and by Bob ($H^{B'}$).
Performing $\calu^{-1}$ reveals the exact origin of those ancillas.}.
Equation~\ref{eqnB+-On01} can be used to describe the interferometer effect on a
general qubit, shown in Equation~\eqref{eqn:interf_evu}.

The states sent by Alice during the ``$xy$-BB84'' protocol
evolve in the interferometer as follows:
\begin{eqnarray}
\begin{array}{rcl}
\label{eqnB+-On+-}
\ket{0_x}_{A}  &\mapm{\phi=0}  &
 (\fet{s_0}_{B}  \phantom{{}-2\fet{s_1}_B} -\fet{s_2}_{B}
 +i\fet{d_0}_{B} +2i \fet{d_1}_{B} +i\fet{d_2}_{B}) \thickspace / \sqrt{8}\\
\ket{1_x}_{A}  &\mapm{\phi=0} &
 (\fet{s_0}_{B} -2 \fet{s_1}_{B} +\fet{s_2}_{B}
+i\fet{d_0}_{B} \phantom{{}+2i\fet{d_1}_B} -i\ket{d_2}_{B}) \thickspace / \sqrt{8}  \\
\ket{0_y}_{A}  &\mapm{\phi=\pi/2}&
 (\fet{s_0}_{B} \phantom{{}+2i\fet{s_1}_B} +\fet{s_2}_{B}
 +i\fet{d_0}_{B} -2 \fet{d_1}_{B} -i\fet{d_2}_{B}) \; /\sqrt{8}\\
\ket{1_y}_{A}  & \mapm{\phi=\pi/2} &
 (\fet{s_0}_{B}  {}-2i \fet{s_1}_{B} -\fet{s_2}_{B}
+i\fet{d_0}_{B} \phantom{{}+2\fet{d_1}_B} +i\fet{d_2}_{B}) \; /\sqrt{8}
\end{array}
\end{eqnarray}
Bob can distinguish the computation basis elements
of bases $x$ and $y$, measuring time-bin $t_1$, i.e.\ the states $\fet{d_1}$ for $\ket{0}$
and $\fet{s_1}$ for $\ket{1}$ in the measured basis. 
Other states give Bob no information
about the state sent by Alice.


\subsection{$H^{B^{-1}}$ of the ``$xyz$-six-state'' scheme}
\label{app:interf_HB-1_basis}
Let Bob be using interferometric
setups $\calu_{B_x}$ and measuring 6 modes 
(corresponding the space with a basis
state $\fet{n_{s_0}n_{s_1}n_{s_2}n_{d_0}n_{d_1}n_{d_2}}_B$) with
one or less photons.
Following Definition~\ref{def:HB-1+anc},
the states spanning the space $H^{B^{-1}}$ can be derived
using Equation~\eqref{eqn:interf_evu} (adjusted to the appropriate space):
\begin{align}
\nonumber
\fet{000000}_B &\mapm{\calu_{B_x}^{-1} } \fet{00000000}_{PB'}
\\
\nonumber
\fet{010000}_B &\mapm{\calu_{B_x}^{-1} }
    \frac{1}{2} \left (-\fet{01000000}_{PB'} + \fet{00100000}_{PB'}
    -i \fet{00000100}_{PB'} -i\fet{00000010}_{PB'}    \right )
\\
\nonumber
\fet{000010}_B &\mapm{\calu_{B_x}^{-1}}
    \frac{1}{2} \left (-i\fet{01000000}_{PB'} -i \fet{00100000}_{PB'}
    + \fet{00000100}_{PB'} -\fet{00000010}_{PB'}    \right )
\\
\nonumber
\fet{000000}_B &\mapm{\calu_{B_z}^{-1} } \fet{00000000}_{PB'}
\\
\nonumber
\fet{001000}_B &\mapm{\calu_{B_z}^{-1}}
    \frac{1}{2} \left (-\fet{00100000}_{PB'} + \fet{00010000}_{PB'}
    -i \fet{00000010}_{PB'} -i\fet{00000001}_{PB'}    \right )
\\
\nonumber
\fet{000100}_B &\mapm{\calu_{B_z}^{-1}}
    \frac{1}{2} \left (-i\fet{10000000}_{PB'} -i \fet{01000000}_{PB'}
    + \fet{00001000}_{PB'} -\fet{00000100}_{PB'}    \right )
\\
\nonumber
\fet{000000}_B &\mapm{\calu_{B_y}^{-1} } \fet{00000000}_{PB'}
\\
\nonumber
\fet{010000}_B &\mapm{\calu_{B_y}^{-1} }
    \frac{1}{2} \left (i\fet{01000000}_{PB'} + \fet{00100000}_{PB'}
    - \fet{00000100}_{PB'} -i\fet{00000010}_{PB'}    \right )
\\
\fet{000010}_B &\mapm{\calu_{B_y}^{-1}}
    \frac{1}{2} \left (-\fet{01000000}_{PB'} -i \fet{00100000}_{PB'}
    -i \fet{00000100}_{PB'} -\fet{00000010}_{PB'}    \right )
\end{align}
defined over the space $H^P \otimes H^{B'}$ with basis state
$\fet{a_{t'_{-1}}a_{t'_0}a_{t'_1}a_{t'_2}b_{t'_{-1}}b_{t'_0}b_{t'_1}b_{t'_2}}_{PB'}$.
Note that performing  $\calu^{-1}$ requires an additional ancilla, 
since the modes number increases from six to eight.


\subsection{QSoP of the ``$xy$-BB84'' scheme}
\label{app:QSoP2ModesIntf}

 Assume Bob measures only time-bin $t_1$ in both output arms of the
 interferometer, i.e.\ the measured space is $H^B$ subspace
 spanned by $\fet{0,n_{s_1},0,0,n_{d_1},0}_B$. 
 Assuming a single-photon restriction,
 the reversed space, of that measured space that is spanned by:
 \begin{align}
 \nonumber
 \fet{000000}_{B} &\mapm{\calu_{B_x}^{-1}}  \fet{0000}_{PB'}
 \\
 \nonumber
 \fet{010000}_{B} &\mapm{\calu_{B_x}^{-1}}
    \frac{1}{2} \left (-\fet{1000}_{PB'} + \fet{0100}_{PB'}
    -i \fet{0010}_{PB'} -i\fet{0001}_{PB'} \right )
 \\
 \nonumber
 \fet{000010}_{B} &\mapm{\calu_{B_x}^{-1}}
     \frac{1}{2} \left (-i\fet{1000}_{PB'} -i \fet{0100}_{PB'}
    + \fet{0010}_{PB'} -\fet{0001}_{PB'}   \right )
 \\
 \nonumber
 \fet{000000}_{B} &\mapm{\calu_{B_y}^{-1}} \fet{0000}_{PB'}
 \\
 \nonumber 
 \fet{010000}_{B} &\mapm{\calu_{B_y}^{-1}}
     \frac{1}{2} \left (i\ket{1000}_{PB'} + \fet{0100}_{PB'}
     - \fet{0010}_{PB'} -i\fet{0001}_{PB'} \right )
 \\
 \fet{000010}_{B} &\mapm{\calu_{B_y}^{-1}}
     \frac{1}{2} \left (-\fet{1000}_{PB'} -i \fet{0100}_{PB'}
     -i \fet{0010}_{PB'} -\fet{0001}_{PB'} \right )
 \label{eqn:x4bb84_HB-1_basis}
 \end{align}
 as can be verified using Equation~\eqref{eqn:interf_evu}.
 The space $H^{B^{-1}}$ is embedded in a 4-mode space $H^P\otimes H^{B'}$,
 having the basis
 element  $\fet{a_{t'_0}a_{t'_1}b_{t'_0}b_{t'_1}}_{PB'}$, i.e.\
 Alice modes at times $t'_0$ and $t'_1$ and Bob's added ancillary modes at times
 $t'_0$ and $t'_1$ respectively.
 The resulting six states \eqref{eqn:x4bb84_HB-1_basis}
 span a 4-dimensional space, i.e.\ $H^{B^{-1}}=H_4$. The QSoP in this special case
 is $H^P = H_3$, spanned by $\fet{a_{t'_0}a_{t'_1}}$ with one or less photons.
\end{appendix}

\end{document}